\documentclass[11pt]{article}
\usepackage{graphicx}
\usepackage{epstopdf}
\usepackage{amsmath}
\usepackage{amssymb}
\usepackage{amsfonts}
\usepackage{amsthm}
\usepackage[usenames]{color}
\usepackage{array}
\usepackage{tabularx}
\usepackage{booktabs}           
\usepackage{caption}
\usepackage[usenames,dvipsnames,svgnames,table]{xcolor}
\usepackage[utf8]{inputenc}
\usepackage{lmodern}
\usepackage{verbatim}
\usepackage{amsmath}
\usepackage{amsfonts}
\usepackage{amssymb}
\usepackage{eso-pic}
\usepackage{graphicx}
\usepackage[final]{pdfpages}
\usepackage{pdfpages}
\usepackage[usenames,dvipsnames,svgnames,table]{xcolor}
\numberwithin{equation}{section}
\newcommand{\ap}{\ensuremath{\alpha'}}
\usepackage[
      colorlinks=true,
      linkcolor=blue,
      urlcolor=blue,
      filecolor=blue,
      citecolor=red,
      pdfstartview=FitV,
      pdftitle={},
      pdfauthor={},
      pdfsubject={},
      pdfkeywords={},
      pdfpagemode={},
      bookmarksopen=true
]{hyperref}

\usepackage{amsmath}
\usepackage{empheq}
\usepackage{mathtools}
\usepackage{epsfig}

\def\beq{\begin{eqnarray}}
\def\eeq{\end{eqnarray}}

\def\g{\gamma}

\def\e{\epsilon}

\def\k{\kappa}

\def\be{\begin{equation}}
\def\ee{\end{equation}}
\def\bea{\begin{eqnarray}}
\def\eea{\end{eqnarray}}

\newcommand{\rom}[1]{\mathrm{#1}}

\def\cF{\mathcal{F}}

\def\nn{\nonumber}

\definecolor{X}{rgb}{0,0,1}
\definecolor{Y}{rgb}{1,0,0}
\definecolor{Z}{rgb}{0,51,0}

\newcommand{\del}{\nabla}

\numberwithin{equation}{section}

\renewcommand{\thefootnote}{\fnsymbol{footnote}}

\begin{document}

\begin{centering}

\thispagestyle{empty}

{\flushright {Preprint  NISER \& CMI 2018}\\[15mm]}

{\LARGE \textsc{A Generalised Garfinkle-Vachaspati Transform}} \\

 \vspace{0.8cm}

{\large 
Deepali Mishra$^{1,4}$, Yogesh K.~Srivastava$^{1,4}$, and  Amitabh Virmani$^{2,3,4 }$}\footnote[1]{Currently on lien from Institute of Physics, Sachivalaya Marg, Bhubaneswar, Odisha, India 751005. }
\vspace{0.5cm}

\begin{minipage}{.9\textwidth}\small  \begin{center}
${}^{1}${National Institute of Science Education and Research (NISER), \\ Bhubaneswar, P.O. Jatni, Khurda, Odisha, India 752050}\\
  \vspace{0.5cm}
$^2$Chennai Mathematical Institute, H1 SIPCOT IT Park, \\ Kelambakkam, Tamil Nadu, India 603103\\
  \vspace{0.5cm}
$^3$Institute of Physics, Sachivalaya Marg, \\ Bhubaneswar, Odisha, India 751005 \\
  \vspace{0.5cm}
$^4$Homi Bhabha National Institute, Training School Complex, \\ Anushakti Nagar, Mumbai India 400085 \\
  \vspace{0.5cm}
{\tt deepalimishra, yogeshs@niser.ac.in, avirmani@cmi.ac.in}
\\ $ \, $ \\

\end{center}
\end{minipage}

\end{centering}

\renewcommand{\thefootnote}{\arabic{footnote}}

\begin{abstract}
The Garfinkle-Vachaspati transform is a deformation of a metric in terms of a null, hypersurface orthogonal, 
Killing vector $k^\mu$. We explore a generalisation of this deformation in type IIB supergravity taking motivation from certain studies of the D1-D5 system. We consider solutions of minimal six-dimensional supergravity admitting null Killing vector $k^\mu$  trivially lifted to type IIB supergravity by the addition of  four-torus directions. The torus directions provide covariantly constant spacelike vectors $l^\mu$.  We show that the original  solution can be deformed  as
$
g_{\mu \nu} \to g_{\mu \nu} + 2 \, \Phi k_{(\mu}l_{\nu)},  \  C_{\mu \nu} \to C_{\mu \nu} - 2 \, \Phi k_{[\mu}l_{\nu]}, 
$
provided the two-form supporting the original spacetime satisfies $i_k (dC) = - d k$, and
where $\Phi$ satisfies the equation of  a minimal massless scalar field on the original spacetime.
We show that the  condition $i_k (dC) = - d k$ is  satisfied by all supersymmetric solutions admitting null Killing vector. Hence all supersymmetric solutions  of minimal six-dimensional supergravity can be deformed via this method. As an example of our approach, we work out the deformation on a class of D1-D5-P geometries with orbifolds. We show that the deformed spacetimes are smooth and identify their CFT description. Using Bena-Warner formalism, we also express the deformed solutions in other duality frames.
\end{abstract}

\newpage
\tableofcontents

\setcounter{equation}{0}

\section{Introduction}

Understanding the entropy of black holes has been a long-standing problem in quantum gravity. In string theory, considerable progress has been made in explaining the entropy of 
black holes in terms of statistical mechanical counting of microstates \cite{Sen:1995in, Strominger:1996sh, David:2002wn,Mathur:2005ai}. For some supersymmetric black holes even exact counting formulae are known \cite{Sen:2007qy, Mandal:2010cj, Dabholkar:2012zz}.  Typically these calculations involve counting states in a string theory system of branes at small coupling and then matching it with the exponential of the Bekenstein-Hawking entropy (or its generalizations like Wald entropy or Sen's quantum entropy function). The success of these calculations give us confidence that string theory has the right ingredients to 
describe black hole entropy. However, unfortunately, these calculations do not tell us how these microstates are to be described in the regime of parameters where we actually have a black hole. 

In the last fifteen years or so, considerable effort has gone in describing microstates of black holes under the fuzzball paradigm \cite{Mathur:2005zp, Bena:2007kg, Skenderis:2008qn, Chowdhury:2010ct}. Various techniques have been developed to construct ``microstate geometries'' --  horizonless, non-singular solutions in supergravity. These solutions are expected to be supergravity approximation to string theory configurations for black hole microstates. The program of constructing such solutions in supergravity has had most success for supersymmetric black holes.  An important step in this program  was the development of general formalisms for classification of 
supersymmetric solutions using Killing spinor techniques. Such a classification was first carried out for minimal  N=2 theory in 4D \cite{Tod:1983pm}, and almost twenty years later for supergravity theories in 5D \cite{Gauntlett:2002nw, Gauntlett:2004qy, Bena:2005va, Berglund:2005vb} and 6D \cite{GMR}. The 6D case considered by Gutowski, Martelli and Reall (GMR) is of special interest to us in this work, where the general supersymmetric solution is given in terms of a 2D fiber over a 4D almost hyper-K\"ahler base space. This form of the 6D solution reduces the problem of solving supergravity equations to a more tractable problem of solving a reduced 
set of equations on the 4D  base space.

In constructing new solutions of supergravity equations, it is also useful to have solution generating techniques. Such techniques   allow us to construct 
new solutions from the known ones. A useful solution-generating technique is the  Garfinkle-Vachaspati transform \cite{Garfinkle:1990jq}. It goes as follows:  given a spacetime configuration with metric $g_{\mu \nu }$ admitting a null, Killing, and hypersurface orthogonal vector field $k^\mu$, i.e., satisfying the following properties,
\begin{align}
  \label{eq:2}
  k^\mu k_\mu &= 0, &
    \del_{(\mu} k_{\nu)}  &=0, &   \del_{[\mu} k_{\nu]}  &= k_{[\mu}\del_{\nu]} S,
\end{align}
for some scalar function $S$,  one can construct a new exact solution
of the equations of motion as,
\begin{equation}
  \label{eq:3}
g'_{\mu \nu}= g_{\mu\nu} + e^{-S} \, \chi \, k_{\mu }k_{ \nu}.
\end{equation}
The new metric $g'_{\mu \nu}$ describes a gravitational wave on the
background $g_{\mu \nu}$ provided the matter fields, if
any, satisfy some conditions \cite{Kaloper:1996hr} and the function $\chi$ satisfies
\begin{align}
\Box \chi &= 0, & k^{\mu}\partial_{\mu}\chi &=0.
\end{align}
This technique has been applied in varied contexts, see e.g.,~\cite{Dabholkar:1995nc, Horowitz:1996th, Banados:1999tw, Hubeny:2003ug, Balasubramanian:2010ys}.

A generalisation of the above Garfinkle-Vachaspati transform was speculated by Lunin, Mathur and Turton (LMT) in \cite{Lunin:2012gp}. Motivated by previous work of Mathur and Turton \cite{Mathur:2011gz, Mathur:2012tj}, LMT considered  supersymmetric deformations of GMR solutions lifted to ten dimensions that add travelling waves. They noticed that the deformed solutions can be written as a generalisation of the Garfinkle-Vachaspati  
 transform, i.e.,
\bea
g'_{\mu \nu} &=& g_{\mu \nu} + 2 \, \Phi \, k_{(\mu} l_{\nu)}, \label{GV1:intro}  \\
  C'_{\mu \nu} &=& C_{\mu \nu} - 2 \, \Phi \, k_{[\mu}l_{\nu]},
\label{GV2:intro}
\eea
where $k^\mu$ is a null, Killing, but need not be hypersurface orthogonal, and  $l^\mu$ is a covariantly constant unit normalised spacelike vector, and $\Box \Phi = 0$.

The difference from the usual Garfinkle-Vachaspati transform comes due to the presence of spacelike Killing vector $l^{\mu}$ and additional two-form potential $C_{\mu \nu}$. In addition, 
the hypersurface orthogonality condition for the  null Killing vector $k^{\nu}$ is not required.  A main aim of  this paper is to present a derivation of the generalised Garfinkle-Vachaspati solution generating technique \eqref{GV1:intro}--\eqref{GV2:intro} and explore its applications. In particular, we achieve three things:
\begin{enumerate}
\item We show that the generalized Garfinkle-Vachaspati transform \eqref{GV1:intro}--\eqref{GV2:intro} is a solution generating technique for ten-dimensional IIB theory.  We show that given a solution of minimal six-dimensional supergravity admitting a null Killing vector, and satisfying the condition
\be
k^\mu F_{\mu \nu \rho} = -(\nabla_\nu k_\rho -\nabla_\rho k_\nu), \label{cond}
\ee 
we can 
get another solution of type IIB theory.  As long as  condition \eqref{cond} is satisfied, we do not require supersymmetry.
The technique allows to add wave-like deformations. 

\item We give explicit examples of applications of this technique.  We add travelling wave deformations on multi-wound round supertubes and on a class of D1-D5-P backgrounds, generalising examples considered in \cite{Lunin:2012gp}. We pick these examples as their dual CFT interpretations are well understood.  We also present CFT interpretation of the deformed solutions. 

\item  For a class of supersymmetric solutions, we convert from GMR notation to Bena-Warner (BW) notation and using 
string theory dualities present the generalised Garfinkle-Vachaspati transformation in various other duality frames. 
\end{enumerate}

The rest of the paper is organised as follows. In section \ref{sec:GV} we present the generalised Garfinkle-Vachaspati transform as a novel solution generating technique. Details on the proof are presented in appendix \ref{sec:GV_details}. In section \ref{sec:GV} we  compare and contrast the generalised Garfinkle-Vachaspati transform with the original Garfinkle-Vachaspati transform and show that all solutions in the GMR form trivially lifted to ten-dimensions can be deformed via this technique. In section \ref{sec:supertube} we work out the deformation on the multi-wound D1-D5 round supertube and on a class of D1-D5-P backgrounds. In section 
\ref{sec:smoothness} we show that the deformations we add preserve smoothness of the solutions and analyse various global properties of the deformed solutions.  In section \ref{sec:CFT} we identity the CFT states for the deformed solutions. In section \ref{sec:dualities} applications of the generalised Garfinkle-Vachaspati transform in different duality frames are explored. Some calculations details from section \ref{sec:dualities} are relegated to appendix \ref{sec:dualities_details}, where a dictionary between the GMR form and the BW form is also worked out. We close with a brief discussion of open problems in the section \ref{sec:disc}.

\section{A generalised Garfinkle-Vachaspati transform}
\label{sec:GV}

In this section, we present the generalised Garfinkle-Vachaspati transform as a novel solution generating technique. The technique allows to add wave-like deformations on solutions of minimal six-dimensional supergravity embedded in ten-dimensional IIB theory. 

We establish that the generalised Garfinkle-Vachaspati transform, 
\bea
g'_{\mu\nu}&=&g_{\mu\nu}+\Phi (k_{\mu}l_{\nu}+k_\nu l_\mu),\\ \label{transformed_metric:main_text}
 C_{\mu \nu}' &=& C_{\mu \nu} - \Phi ( k_{\mu}l_{\nu} - l_{\mu} k_{\nu}),
\label{transformed_C:main_text}
\eea
is a valid solution generating technique via a direct calculation. We show that the left and the right hand side of the Einstein equations transform in the exactly the same way, thereby establishing that if we start with a solution, we can deform it to a new solution. In our convention, Einstein equations are, 
\begin{equation}
R_{\mu\nu} =\frac{1}{4}F_{\mu\lambda\sigma}{F_\nu}^{\lambda\sigma} \label{einstein_eqs:main_text},
\end{equation}
together with $F_{\mu\lambda\sigma} F^{\mu\lambda\sigma} = 0, F_{\mu\lambda\sigma} = (d C)_{\mu\lambda\sigma}$ and 
matter field equations are, 
\be
\del_\mu F^{\mu \nu \rho} =0. 
\ee
The vector $k^\mu$ appearing in  \eqref{transformed_metric:main_text} is a null Killing vector.
The vector $l^\mu$ appearing in  \eqref{transformed_metric:main_text} is a unit normalised covariantly constant spacelike (Killing) vector orthogonal to $k^\mu$, and  $\Phi$ is a massless scalar on the original background spacetime $g_{\mu\nu}$,
\be
\Box\Phi =0,  \label{massless_scalar:main_text}
\ee
compatible with the Killing symmetries, i.e., $k^\mu \nabla_\mu \Phi =0$ and $l^\mu \nabla_\mu \Phi =0$. The transformed configuration also has $k^\mu$ and $l^\mu$ as  Killing symmetries.

We present the details of the calculation of deformations of the left and the right hand side of Einstein equations in appendix \ref{sec:GV_details}. Here we simply note that the left hand side transforms as,
\bea
R'_{\lambda\nu} 
 &=& R_{\lambda\nu}-l_\lambda[ k^\mu (\del_\nu\del_\mu \Phi)+\Phi \square k_\nu ]-l_\nu [ k^\mu (\del_\lambda\del_\mu \Phi)+\Phi \square k_\lambda ]\nonumber \\
&&+\frac{1}{2}(\del_\rho\Phi)(\del^\rho\Phi)k_\lambda k_\nu-\Phi^2(\del_\mu k^\rho)(\del_\rho k^\mu ) l_\lambda l_\nu,
\eea
while the right hand side transforms in the same way as long as, 
\be
i_k (dC) = -dk. \label{transversality:main_text}
\ee 
In appendix \ref{sec:GV_details} we also show that the $3$-form field equation transforms covariantly, i.e.,
\be
\del_\mu F^{\mu \nu \rho} =0 \implies \del'_\mu F'^{\mu\nu\rho}= 0.
\ee

Often in string theory applications there are more than one 
covariantly constant spacelike (Killing) vectors $l^\mu_{(a)}$  orthogonal to $k^\mu$ are available. In such situations, the generalised Garfinkle-Vachaspati transformation technique admits a further generalisation
 \bea
g'_{\mu\nu}&=&g_{\mu\nu}+\sum_{a}\Phi_{(a)} (k_{\mu}l^{(a)}_{\nu}+k_\nu l^{(a)}_\mu),\\
 C_{\mu \nu}' &=& C_{\mu \nu} - \sum_{a}\Phi_{(a)} ( k_{\mu}l^{(a)}_{\nu} - l^{(a)}_{\mu} k_{\nu}) ,
\eea
where $\Phi_{(a)}$ are scalars on the original background spacetime $g_{\mu\nu}$ satisfying $\Box\Phi_{(a)}=0.$

\subsection{Comparison to Garfinkle-Vachaspati transform}
\label{sec:oldGV}
Compared to the Garfinkle-Vachaspati (GV) transform, our solution-generating technique is more restrictive in some ways. 
As shown in \cite{Kaloper:1996hr}, for the GV technique to work the original matter fields have to satisfy certain algebraic transversality conditions. As long as those conditions are satisfied, the matter fields do not transform.
Unlike the GV technique, in our technique the matter fields do transform. There is no uniform prescription  for the transformation of  all matter fields.  We need to do a case by case analysis.  For the two-form gauge field considered in this paper, the transformation is \eqref{transformed_C:main_text}, provided the untransformed $3$-form field strength satisfies the differential transversality condition \eqref{transversality:main_text}. The differential transversality  condition \eqref{transversality:main_text} is analogous to the transversality condition for the GV technique, though now it is  a differential condition rather than an algebraic condition.

In the next subsection we show that the differential transversality  condition \eqref{transversality:main_text} is satisfied for all supersymmetric solutions written in the GMR form. However, to the best of our understanding, conditions for 
 having supersymmetric solutions are more extensive than just the above differential transversality condition. We suspect that our 
 solution-generating technique finds applications in non-supersymmetric settings as well, provided the differential transversality condition \eqref{transversality:main_text} is satisfied, though  we do not work out any non-supersymmetric example in this paper.

The differential transversality  condition is consistent with Einstein equations. To see this,  contract equations \eqref{einstein_eqs:main_text} with the $k^\mu k^\nu$ as:
 \begin{equation}
R_{\mu\nu} k^\mu k^\nu=\frac{1}{4}k^\mu F_{\mu\lambda\sigma}k^\nu {F_\nu}^{\lambda\sigma} \label{einstein1_eqs},
\end{equation}
From the fact that $k^{\mu}$ is a Killing vector, we have the identity \be
k^\lambda \square 
k_\lambda = - R_{\lambda\rho}k^\lambda k^\rho.
\ee 
From this, it follows that
\bea
R_{\lambda\rho}k^\lambda k^\rho &=& - k^\lambda \square k_\lambda \\
&=& - \left( \nabla^{\mu}(k^\lambda \nabla_\mu k_\lambda) - (\nabla^\mu k^\lambda)(\nabla_\mu k_\lambda) \right) \\
&=& \frac{1}{4}\left[(\nabla^\mu k^\lambda - \nabla^\lambda k^\mu)(\nabla_\mu k_\lambda - \nabla_\lambda k_\mu)\right],
\eea
where we have used the fact that $k^{\mu}$ is null and Killing. Equating this with the right hand side of equation \eqref{einstein1_eqs}, we have 
\be
k^\mu F_{\mu\lambda\sigma}k^\nu {F_\nu}^{\lambda\sigma}= (\nabla^\lambda k^\sigma - \nabla^\sigma k^\lambda)(\nabla_\lambda k_\sigma - \nabla_\sigma k_\lambda),
\ee
which is  the ``square" of this differential transversality condition \eqref{transversality:main_text}. 

\subsection{Application to supersymmetric solutions}
\label{sec:GMR}
We can now apply the generalized Garfinkle-Vachaspati transform to supersymmetric solutions of minimal six-dimensional supergravity. For this set-up, our results are the same as \cite{Lunin:2012gp}, so we shall be brief. In that reference, the authors showed that supersymmetric solutions of minimal six-dimensional supergravity embedded in ten-dimensional IIB theory can be deformed. They showed consistency with Einstein equations by showing that the deformed solutions are supersymmetric solutions  of ten-dimensional IIB theory.  The arguments presented there are of very different nature compared to the direct derivation of the generalized GV transform presented in this work. We now show the connection.

Supersymmetric solutions of minimal six-dimensional supergravity, trivially lifted to ten dimensions, can be written as \cite{GMR, Lunin:2012gp}
\be
ds^2= -H^{-1}(dv + \beta)\Bigl(du + \omega + \frac{\cF}{2}(dv + \beta)\Bigr) + H h_{mn} dx^m dx^n + dz_i dz_i.
\label{general6d}
\ee
with 
\be
k = \frac{\partial}{\partial u},
\ee 
being the null Killing vector. To apply the generalized GV transform, we can pick any one of the spacelike covariantly constant (Killing) vector provided by the torus directions. We pick, say,
 \be 
l = \frac{\partial}{\partial z_4}.
\ee 
For the successful application of the generalized GV transform, we only need to check that the field strength supporting \eqref{general6d} satisfies the differential transversality condition \eqref{transversality:main_text}. The Killing spinor equation implies this differential transversality  condition \cite{GMR}. We can also explicitly check that it is satisfied using appendix A of \cite{Lunin:2012gp}. To this end, consider $k^\mu F_{\mu \nu \rho}$ :
\bea
k^\mu F_{\mu \nu \rho} &=&  F_{u \nu \rho} \\
&=& \partial_u C_{\nu  \rho}  + \partial_\rho C_{u \nu} + \partial_\nu C_{\rho u} \\
&=& - (\partial_\nu C_{u\rho} -\partial_\rho C_{u \nu} ).
\eea
We see that the differential transversality condition is equivalent  to showing $C_{u\nu} = k_\nu$, upto possible gauge transformations. Looking at the equation (A.6) of \cite{Lunin:2012gp}, we see that indeed it is the case for the general GMR solution:
\be
C_{u\nu}dx^\nu = -\frac{1}{2H}(dv +\beta)= k_\nu dx^\nu.
\ee

\section{Deformation of a class of D1-D5-P backgrounds}
\label{sec:supertube}

 In this section we present explicit examples of our general construction. We consider two classes of examples: multi-wound D1-D5 round supertubes and a class of D1-D5-P backgrounds. Throughout this section, $Q_1=Q_5=Q$, 
where 
\be \label{eq:Q} 
Q_1~=~ \frac{g \ap^3}{V} n_1 \,, \qquad Q_5 ~=~  g \ap n_5 \,, \qquad (2\pi)^4 V ~=~ \mbox{vol}(T^4).  
\ee

Multi-wound D1-D5 round supertubes were constructed in~\cite{Balasubramanian:2000rt, Maldacena:2000dr}. This family is parametrised by an integer $k$ via,
\be
\gamma=\frac{1}{k}, \qquad \qquad k = 1, 2 \ldots, N, \qquad \qquad N = n_1 n_5.
\ee
The case $k=1$ corresponds to singly wound D1-D5 supertube. This configuration is dual to Ramond vaccum $|0\rangle_\rom{R}$. The $k\neq1$ members of the family are obtained by acting with certain twist operator such that the resulting states have $N/k$ component strings \cite{Lunin:2001jy}. For $k\neq1$ the geometries have conical singularities. The metric takes the form,
\bea
ds^2_0 & = & -\frac{1}{h} (dt^2-dy^2) + h f \left( \frac{dr^2}{r^2 +
a^2\gamma^2} + d\theta^2 \right)
\nonumber \\
         &+& h \Bigl( r^2 +
\frac{a^2\gamma^2\,Q^2\,\cos^2\theta}{h^2 f^2} \Bigr)
\cos^2\theta d\psi^2  \nonumber \\
&+& h\Bigl( r^2 + a^2\gamma^2 -
\frac{a^2\gamma^2\,Q^2 \,\sin^2\theta}{h^{2} f^{2} }
\Bigr) \sin^2\theta d\phi^2  \nonumber \\
&-& \frac{2a\gamma\,Q }{hf}
(\cos^2\theta \,dy\,d\psi + \sin^2\theta \,dt\,d\phi)
+
dz_i dz_i, \label{orbifold}
\eea
and the two-form field takes the form,
\begin{align}
C^{0}_{ty}&=-\frac{Q}{Q+f}, &
C^{0}_{t\psi} &= -\frac{Qa \gamma \cos^2\theta}{Q+f}, &  \nonumber\\
C^{0}_{y\phi}&=-\frac{Qa \gamma \sin^2\theta}{Q+f}, & 
C^{0}_{\phi\psi} &= Q\cos^2\theta+\frac{Qa^2 \gamma^2 \sin^2\theta\cos^2\theta}{Q+f}, &
\end{align}
where
\begin{align}
f&=r^2+a^2 \gamma^2 \cos^2\theta, & h&=1+\frac{Q}{f}.
\end{align}
The $y$ coordinate is periodic with periodicity $2 \pi R_y$, and the parameter $a$ is related to the size $R_y$ of the $y$-circle as,
\be
  a=\frac{Q}{R_y} . \label{a_eq}
\ee

In the large $R_y$ limit, the above geometry has a long $AdS_3 \times S^3 \times  T^4$ throat.  The throat together with the cap region is described by the metric obtained by focusing  on the region of the spacetime with $ r \ll \sqrt{Q}$. In this limit the metric becomes locally $AdS_3 \times S^3$ with a $\mathbb{Z}_k$ orbifold at $r=0,~\theta = \frac{\pi}{2}$.

Linear deformation of the type obtained via our Garfinkle-Vachaspati transform on this solution were studied in \cite{Mathur:2012tj}. We proceed by writing the linear perturbation from reference \cite{Mathur:2012tj} in a suggestive form.  We will then see that the deformation is valid non-linearly. To begin with, let us start by writing the  
background solution in  GMR form \eqref{general6d}:
\bea
\label{GMR_form}
ds^2_0&=&-\frac{1}{h}\left[du+A\right]\left[dv+B\right]+
h ds^2_\rom{base}
+dz_i dz_i, \label{bg_1} \\
C_0&=&\frac{1}{2h}[dv+B]\wedge [du+A]+Q \frac{(r^2 +a^2 \gamma^2)}{f}\, c_\theta^2 \, d\phi\wedge d\psi, \label{bg_2} 
\eea
with
\be
ds^2_\rom{base} = \frac{f}{r^2+a^2 \gamma^2}dr^2+d\theta^2+r^2c_\theta^2 d\psi^2+(r^2+a^2 \gamma^2)s_\theta^2d\phi^2,
\ee
and one-forms
\bea
A&=&\frac{a\gamma Q}{f}\{s_\theta^2 d\phi-c_\theta^2 d\psi\}, \\
B&=&\frac{a\gamma Q}{f}\{s_\theta^2 d\phi+c_\theta^2 d\psi\},
\eea
where $c_\theta = \cos \theta$ and $s_\theta = \sin \theta$. 

The linear perturbation in reference~\cite{Mathur:2012tj} was constructed in the gauge
\be
h_{\mu z}+(C-C_0)_{\mu z}=0,
\ee
where $z$ is one of the four-torus coordinates. 
The explicit form of the solution with added linear perturbation is 
\be
ds^2 = ds^2_0 + 2 \, \e \, e^{-in\frac{v}{R_y}} \left(\frac{r^2}{r^2+a^2\gamma^2}\right)^{\frac{nk}{2}} K \ dz, 
\ee
\be%
C = C_0 + \e \, e^{-in\frac{v}{R_y}}
\left(\frac{r^2}{r^2+a^2\gamma^2}\right)^{\frac{nk}{2}} \ dz \ \wedge K 
\label{C2_new_K}, 
\ee
where
\be
K = 
\frac{Q}{Q+f}\left[
dv-a \gamma (c_\theta^2 d\psi+s_\theta^2 d\phi)\right]+
\frac{ia \gamma Q}{r(r^2+a^2 \gamma^2)}dr. \label{K_def}
\ee
We can  simplify  this form of the solution by adding a pure-gauge piece. We start by observing that $K$ defined in \eqref{K_def} can also be written as
\be
K = 
-\frac{f}{Q+f}\left[
dv+B\right]+ dv + 
\frac{ia \gamma Q}{r(r^2+a^2 \gamma^2)}dr.  \label{K_def_2}
\ee
Contribution to $C$, cf.~\eqref{C2_new_K}, from the last two terms of $K$ in the form of equation \eqref{K_def_2} can be identified as a complete differential
\bea
e^{-in\frac{v}{R_y}}
\left(\frac{r^2}{r^2+a^2\gamma^2}\right)^{\frac{nk}{2}} \left[ dv + 
\frac{ia \gamma Q}{r(r^2+a^2 \gamma^2)}dr \right] \equiv d \Psi,
\eea
where
\be
\Psi = \frac{i R_y}{n} e^{-in\frac{v}{R_y}} \left(\frac{r^2}{r^2+a^2\gamma^2}\right)^{\frac{nk}{2}}.
\ee
As a result we can gauge away these pieces. Specifically, consider the diffeomorphism and the gauge transformation,
\bea
\xi_z &=& - \Psi, \\ 
 \Lambda &=& \Psi d z.
\eea
The new metric 
\bea
g^\rom{new}_{\mu \nu} & = & g_{\mu \nu} + \e  \ \del_{(\mu} \xi_{\nu)},
\eea
takes the form
\bea
ds^2_\rom{new} &=& g^\rom{new}_{\mu \nu}  dx^\mu d x^\nu \\ 
&=& ds^2_0 + 2 \, \e \, e^{-in\frac{v}{R_y}} \left(\frac{r^2}{r^2+a^2\gamma^2}\right)^{\frac{nk}{2}} \left\{ -\frac{f}{Q+f}\left[
dv+B\right] \right\} dz , \label{g_new} 
\eea
and new two-form field is
\bea
C_\rom{new} &= & C + \e \ d \Lambda \\
  &=& C_0 + \e \, e^{-in\frac{v}{R_y}}
\left(\frac{r^2}{r^2+a^2\gamma^2}\right)^{\frac{nk}{2}}\left\{\frac{f}{Q+f}\left[
dv+B\right] \right\} \wedge dz.  \label{C2_new}
\eea
The configuration \eqref{g_new} and \eqref{C2_new} is a generalised Garfinkle-Vachaspati transform of background \eqref{bg_1}--\eqref{bg_2}. It is 
a non-linear solution of ten-dimensional IIB supergravity. 
Therefore, from now onwards we set $\e = 1.$ 
Realising that $\frac{f}{Q+f}$ is simply $\frac{1}{h}$ we observe that the above solution is compatible with the form \eqref{GMR_form}, provided we shift the one-form $du$ as
\bea
du &\to& du + \Phi \, d z, \\
\Phi &=&
 2 \, \left(\frac{r^2}{r^2+a^2\gamma^2}\right)^{\frac{nk}{2}} e^{-in\frac{v}{R_y}}.\label{Phi_special}
 \eea

The scalar field $\Phi$ satisfies $\Box_0 \Phi =0$ with respect to the background metric $ds_0^2$. This deformation is therefore of the form \eqref{general6d}.
We can generalise the above deformation further.  Instead of working with the specific solution \eqref{Phi_special}, we can consider the most general $u$-independent solution of the wave equation $\Box_0 \Phi =0$ that remains finite everywhere. Such a solution can be written as a superposition
 \be
\Phi = \sum_{n=-\infty}^{\infty} c_n \left(\frac{r^2}{r^2+a^2\gamma^2}\right)^{\frac{|n|k}{2}} e^{-in\frac{v}{R_y}}. \label{scalar_0}
\ee
The requirement that $\Phi$ be real fixes $(c_n)^* = c_{-n}$.

  In fact, we can straightforwardly generalise the above discussion even further.  In references \cite{gms1,gms2} a bigger class of three-charge solutions of IIB supergravity were
constructed that generalise the above backgrounds with one more integer parameter $m$. These solutions are parametrised by parameters $\gamma_1,\gamma_2$ and charges $Q_1$ and $Q_5$. 
The dilaton vanishes for these solutions when the $Q_1$ and $Q_5$ are set equal $(Q_1 = Q_5 = Q)$ and the moduli at infinity are chosen appropriately. In the component string picture of the D1-D5 CFT, these states corresponds $2m+1$ units of spectral flows on the above discussed orbifolds. A more general family is known where the spectral flow parameter is also fractionated \cite{Jejjala:2005yu, Giusto:2012yz, Chakrabarty:2015foa}. For simplicity, we do not consider those states here; we expect our analysis to straightforwardly extend to those cases as well.   The six-dimensional metric  is \cite{gms1,gms2}
\bea
ds^2 & = & -\frac{1}{h} (dt^2-dy^2) + \frac{Q_{p}}{h
f}\left(dt-dy\right)^{2}+ h f \left( \frac{dr^2}{r^2 +
(\g_1+\g_2)^2\eta} + d\theta^2
\right)\nonumber \\
          && + \ h \Bigl( r^2 + \g_1\,(\g_1+\g_2)\,\eta -
\frac{Q^2\,(\g_1^2-\g_2^2)\,\eta\,\cos^2\theta}{h^2 f^2}
\Bigr)
\cos^2\theta d\psi^2  \nonumber \\
&& + \  h\Bigl( r^2 + \g_2\,(\g_1+\g_2)\,\eta +
\frac{Q^2\,(\g_1^2-\g_2^2)\,\eta\,\sin^2\theta}{h^{2} f^{2}
}
\Bigr) \sin^2\theta d\phi^2  \nonumber \\
&& + \ \frac{Q_p\,(\g_1+\g_2)^2\,\eta^2}{h f}
\left( \cos^2\theta d\psi + \sin^2\theta d\phi \right)^{2} \nonumber\\
&& - \ \frac{2 Q }{hf}
\left(\g_1 \cos^2\theta d\psi + \g_2 \sin^2\theta d\phi\right)
(dt-dy)
\nonumber \\
&& - \ \frac{2 Q\,(\g_1+\g_2)\,\eta}{h f}
\left( \cos^2\theta d\psi + \sin^2\theta d\phi \right) dy,
\label{3chargemetric}
\eea
with
\begin{align}
& Q_p =-\g_1\g_2 ,
&\eta &= \frac{Q}{Q + 2 Q_p}, & \\
& f = r^2+ (\g_1+\g_2)\,\eta\, \bigl(\g_1\, \sin^2\theta + \g_2\,\cos^2\theta\bigr), & 
&h =1+\frac{Q}{f} ,& \label{deffh} \\
& \g_1 = - a\,m ,& 
&\g_2  =a\,\Bigl(m+\frac{1}{k}\Bigr). \label{sf}
\end{align}
We consider the range  $m \ge 0, k > 0 \in\mathbb{Z}.$
The two-form field supporting this configuration can be written as \cite{gms2}
\bea
C &=& -\frac{Q c_\theta^2}{Q + f} (\gamma_2 dt + \gamma_1 dy) \wedge d\psi -\frac{Q s_\theta^2}{Q + f} (\gamma_1 dt + \gamma_2 dy) \wedge d\phi \nn \\
& & + \  \frac{(\gamma_1 + \gamma_2) \, \eta \, Q_p}{Q +f} (dt + dy) \wedge (c_\theta^2 d \psi + s_\theta^2 d \phi) - \frac{Q}{Q+f} dt \wedge d y \nn \\
& & - \ \frac{Q c_\theta^2}{Q+f} (r^2  + \gamma_2 (\gamma_1 + \gamma_2) \eta + Q) d\psi \wedge d \phi.
\eea
In this class of metrics when we set $m=0$ we get back to the configuration \eqref{orbifold}. This more general family when written in the GMR form  \eqref{general6d} has quantities $H$, $\cF$, $\beta$, $\omega$  given as~\cite{Giusto:2004kj},
\bea
H&=&h,\\
\cF &=& -\frac{Q_p}{f},\\
 \beta&=& \frac{Q}{f}\,(\g_1 + \g_2)\,\eta\,(\cos^2\theta\,d\psi
+ \sin^2\theta\,d\phi), \\
 \omega&=& \frac{Q}{f}\,\Bigl[\Bigl(2\g_1 -
(\g_1 + \g_2)\,\eta\,\Bigl(1-2 \frac{Q_p}{f}\Bigr)\Bigr)\,\cos^2\theta\,d\psi \nn \\
&&  \quad + \Bigl(2\g_2 - (\g_1 + \g_2)\,\eta\,\Bigl(1-2
\frac{Q_p}{f}\Bigr)\Bigr)\,\sin^2\theta\,d\phi\Bigr],
\label{3chargegmr}
\eea
and the base metric $h_{mn}$ given as,
\bea
ds^2_\rom{base} &=& h_{mn} dx^m dx^n= f\left(\frac{dr^2}{r^2+(\g_1
+ \g_2)^2\, \eta} + d\theta^2\right)\nonumber\\
&&+ \frac{1}{f}\Bigl[[r^4 +
r^2\,(\g_1+\g_2)\,\eta\,(2\g_1 - (\g_1-\g_2) \cos^2\theta) +
(\g_1+\g_2)^2\,\g_1^2\,\eta^2\,\sin^2\theta]\cos^2\theta\,d\psi^2
\nn \\
&&+[r^4 +
r^2\,(\g_1+\g_2)\,\eta\,(2\g_2 + (\g_1-\g_2) \sin^2\theta) +
(\g_1+\g_2)^2 \,\g_2^2 \,\eta^2 \,\cos^2\theta]\sin^2\theta
\,d\phi^2\nn \\
&&-2\g_1 \g_2 \,(\g_1+\g_2)^2\, \eta^2\,\sin^2\theta \cos^2\theta \,d\psi d\phi\Bigr].
\label{3chargebase}
\eea

On this rather complicated configuration one can add a general deformation as,
\bea
du &\to& du + \Phi_i \, d z_i, \\
\Phi_i &=&
 \sum_{n=-\infty}^{\infty}c_n^i \, \left(\frac{r^2}{r^2 \left( 1 +  \frac{2a^2}{Q}m\left(m+\frac{1}{k}\right)\right) + \frac{a^2}{k^2}}\right)^{\frac{|n|k}{2}} e^{-in\frac{v}{R_y}}. \label{scalar_m}
\eea
Indeed $\Box \Phi_i = 0$ with respect to the background metric \eqref{3chargemetric}; the index $i$ refers to the four-torus directions.
Note that when $m=0$, scalar \eqref{scalar_m} reduces to deformation scalar \eqref{scalar_0}; when $k=1$ it reduces to the deformation considered in section 5 of \cite{Lunin:2012gp}. The deformed two-form field is,
\bea
C &=& -\frac{1}{2h} [du +\Phi_i \, d z_i] \ \wedge dv + \frac{(\gamma_1 + \gamma_2) }{h f } \left(\eta Q_p - \frac{Q}{2}\right)  [du +\Phi_i \, d z_i] \wedge (c_\theta^2 d\psi + s_\theta^2 d \phi) \nn \\
& & - \  \frac{Q}{2 h f}  (\gamma_2 - \gamma_1) dv  \wedge (c_\theta^2 d \psi - s_\theta^2 d \phi)   \nn \\
& & - \ \frac{Q}{h f} c_\theta^2 (r^2  + \gamma_2 (\gamma_1 + \gamma_2) \eta + Q) d\psi \wedge d \phi.
\eea

The deformed solution has flat asymptotics, however it is not manifest in the above coordinates. In the next section we find a set of coordinates that makes the asymptotic flatness of the solution manifest and read off the charges of the solution. In the following section we identify the CFT states dual to the deformed spacetimes.

\section{Global properties and smoothness of deformed spacetimes}
In this section we present a discussion on asymptotics,  ADM charges,  smoothness and some other global properties and of the deformed spacetime.  The following discussion is a generalisation of the corresponding discussion in \cite{Lunin:2012gp} of D1-D5-P geometries with $k=1$ to D1-D5-P orbifolds parametrised by integer $k \neq 1$. We write out calculations where our analysis offers a simplification, or  a different perspective, or fixes typos/errors over the corresponding discussion in that reference.
\label{sec:smoothness} 

\subsection{Asymptotics}
To find the map between the deformed spacetime and the CFT states, we need to evaluate charges of the deformed spacetime. We first evaluate the charges in the asymptotically flat setting, and in the next section in the $AdS_3 \times S^3 \times T^4$ setting. We assume that $c_0^i=0$ in \eqref{scalar_m}. A constant term in $\Phi$ can be removed by shifting the $u$-coordinate. However, since $y$ and $z_i$ are periodic coordinates, such a shift does have an effect on the global properties of the solution. For simplicity we do not analyse the constant terms in $\Phi_i$ here, and assume they are set to zero. At infinity metric of the deformed spacetime takes the form
\be
ds^2 = - \left[ du + f_i(v) dz_i \right] dv + dr^2 + r^2 d \Omega_3^2 + dz_i dz_i, \label{asymptotic_deformed}
\ee
where
\be
f_i(v) = \lim_{r\to \infty} \Phi_i(r,v) = \sum_{n \neq 0} c_n^i \left(1 +  \frac{2a^2}{Q}m\left(m+\frac{1}{k}\right)\right)^{-\frac{|n|k}{2}} e^{-in\frac{v}{R_y}}. \label{f_i}
\ee
The diffeomorphism that puts the metric \eqref{asymptotic_deformed} in a standard asymptotically flat form and has the property that the new time-coordinate is single valued is:
\bea
z'_i &=& z_i - \frac{1}{2} \int_0^v f_i( \tilde v) d \tilde v, \label{asym_diff1} \\ 
u' &=&\lambda \left[ u + \frac{1}{4}\int_0^{v} f_i ( \tilde v)f_i( \tilde v) d \tilde v\right],  \label{asym_diff2}  \\
v' &=& \frac{v}{\lambda},\label{asym_diff3} 
\eea
with the value of $\lambda$ is fixed by the requirement that the new time coordinate $t' = \frac{1}{2}(u' + v')$
 is a single valued function under $y \sim y + 2 \pi R_y$. This is achieved as follows:  
\bea
t'(y = 2\pi R_y) - t'(y=0) &=& \lambda \left[ \pi R_y + \frac{1}{8}\int_t^{t-2 \pi R_y} f_i(\tilde v) f_i (\tilde v) d \tilde v\right] - \frac{ \pi R_y}{\lambda} \\
&=&  \pi R_y  \left[ \lambda - \frac{1}{\lambda} \right]+\frac{\lambda}{8}\int_0^{-2 \pi R_y} f_i ( \tilde v) f_i ( \tilde v) d \tilde v \\
&=&
 \pi R_y  \left[ \lambda - \frac{1}{\lambda} \right]-\frac{\lambda}{8}\int_{0}^{2\pi R_y} f_i ( \tilde v) f_i ( \tilde v) d \tilde v, 
 \eea
 where in going from the first step to the second we have used the fact that since $f_i(\tilde v)$ are periodic functions in $\tilde v \sim \tilde v - 2 \pi R_y$, the limit of integration $(t, t-2\pi R_y)$ can be changed to  $(0, -2 \pi R_y)$. In going from the second step to the third step, we have once again used the periodic property of the functions $f_i(\tilde v)$ and converted the limit of integration to $(0, 2 \pi R_y)$.
 This fixes the value of $\lambda$ to be:
\be
\lambda^{-2} = \left[ 1 - \frac{1}{8 \pi R_y} \int_0^{2 \pi R_y} f_i( \tilde v) f_i( \tilde v) d \tilde v\right]. \label{lambda2}
\ee
This expression differs from the one written in equation (4.12) of \cite{Lunin:2012gp}; also the value of the function $f_i(v)$ in \eqref{f_i} is different from equation (6.2) of \cite{Lunin:2012gp} when $k=1$.\footnote{We thank David Turton and Oleg Lunin for a detailed discussion on these points. After their paper was accepted for publication, they also independently realised these typos.}

In new coordinates, the asymptotic metric \eqref{asymptotic_deformed} is 
\be
ds^2 = - (dt')^2 + (dy')^2 + dr^2 + r^2 d \Omega^2_3 + dz'_i dz'_i.
\ee
The $z'_i$ coordinates have the same periodicity as the $z_i$ coordinates. The periodicity of the $y'$ coordinate is 
\bea
y'(y = 2\pi R_y) - y'(y=0) &=& \lambda \left[ \pi R_y + \frac{1}{8}\int_t^{t-2 \pi R_y} f_i(\tilde v) f_i (\tilde v) d \tilde v\right] + \frac{ \pi R_y}{\lambda} \\
&=&  \pi R_y  \left[ \lambda + \frac{1}{\lambda} \right]+\frac{\lambda}{8}\int_0^{-2 \pi R_y} f_i ( \tilde v) f_i ( \tilde v) d \tilde v \\
&=&
 \pi R_y  \left[ \lambda + \frac{1}{\lambda} \right]-\frac{\lambda}{8}\int_{0}^{2\pi R_y} f_i ( \tilde v) f_i ( \tilde v) d \tilde v \\
 &=& \frac{2 \pi R_y} {\lambda}.
 \eea
This implies that the deformed solution has asymptotic radius $y' \sim y' + 2 \pi R$, with 
\be
R =  \frac{R_y}{\lambda}.
\ee

The picture is as follows: deformations of a given state are constructed by introducing functions $\Phi_i$, while keeping $n_1, n_5, m, k$ and asymptotic radius $R$ fixed. In order to work with radius $R$ (as opposed to $R_y$) we introduce
\be
h_i(v') = f_i(v) = f_i(\lambda v').
\ee
and we also note that
\be
\lambda^{-2} = 1 - \frac{1}{8 \pi R} \int_0^{2 \pi R} h_i( \tilde v') h_i( \tilde v') d \tilde v'. \label{lambda}
\ee

\subsection{Charges}
Now that we know the coordinate transformations that bring the metric in the standard flat form asymptotically, we can work out the charges. We extend the diffeomorphism \eqref{asym_diff1}--\eqref{asym_diff3} to finite radial coordinates as:
\bea
z'_i &=& z_i - \frac{1}{2} \int_0^v \Phi_i( \tilde v) d \tilde v,  \label{diff1} \\ 
u' &=&\lambda \left[ u + \frac{1}{4}\int_0^{v} \Phi_i ( \tilde v)\Phi_i( \tilde v) d \tilde v\right], \\
v' &=& \frac{v}{\lambda}. \label{diff3}
\eea
This choice simplifies the extraction of charges. At large values of $r$ we find\footnote{In the following equations, we only write  components of the metric that are relevant for the computation of the gravitational charges. The are other components with $\frac{1}{r^2}$ terms.},
\begin{align}
&g_{t't'} = -1 + \frac{1}{r^2} \left( Q + \lambda^2 Q_p + \frac{1}{4} \lambda^2 Q h_i h_i\right) + \ldots\\
&g_{t'y'}= - \frac{\lambda^2}{r^2} \left( Q_p + \frac{1}{4} Q h_i h_i  \right)+ \ldots\\
&g_{y'y'}= 1+ \frac{1}{r^2} \left( -Q + \lambda^2 Q_p + \frac{1}{4} \lambda^2 Q h_i h_i\right)+ \ldots\\
&g_{t'z_i} = \frac{\lambda Q}{2 r^2} h_i+ \ldots\\
&g_{t'\phi}=- \frac{\lambda Q}{r^2} s_\theta^2 \left( \gamma_2 - \frac{\gamma_1 + \gamma_2}{2} \eta \left( 1- \frac{1}{4}h_i h_i - \frac{1}{\lambda^2}\right)\right)+ \ldots\\
&g_{t'\psi}=-\frac{\lambda Q}{r^2} c_\theta^2 \left( \gamma_1 - \frac{\gamma_1 + \gamma_2}{2} \eta \left( 1- \frac{1}{4}h_i h_i - \frac{1}{\lambda^2}\right)\right)+ \ldots.
\end{align}

From these components we can extract the charges.
The ADM momenta of the solution are  given by
\bea
P_i &=& - \frac{\pi}{4 G_N} \int_0^{2\pi R} dy \ r^2 \ \delta g_{t'z_i} =  0, \\
P_{y'} &=& - \frac{\pi}{4 G_N} \int_0^{2\pi R} dy \ r^2 \ \delta g_{t'y'} =  \frac{\pi \lambda^2}{4G_N} \left( 2 \pi R \ Q_p + \frac{1}{4} Q \int_{0}^{2\pi R}h_i h_i dy'  \right),
\eea
where we have used the fact that $c_0^i = 0$ and where $G_N = \frac{\pi^2 \alpha'^4 g^2}{2 V}$ is the six-dimensional Newton's constant.

The ADM mass is \cite{Harmark:2004ch}
\bea
M &=&  \frac{\pi}{8 G_N} \int_0^{2\pi R} dy \ r^2 \ (3 \delta g_{t't'}  - \delta g_{y'y'}) \\
&=& \frac{\pi}{4G_N} (2 Q) (2 \pi R) +  \frac{\pi \lambda^2}{4G} \left( 2 \pi R \ Q_p + \frac{1}{4} Q \int_{0}^{2\pi R}h_i h_i dy'  \right)\\
&=&  \frac{\pi}{4G_N} (2 Q) (2 \pi R) + P_{y'}.
\eea
Not surprisingly, the BPS bound is saturated; addition of momentum shifts the mass by $P_{y'}$. Using \eqref{eq:Q} can rewrite the ADM momentum $P_{y'}$ as
\be
P_{y'} = \frac{n_1 n_5}{R} \left[ m \left( m + \frac{1}{k} \right)+ \frac{Q}{4a^2} \frac{1}{2\pi R} \int_0^{2\pi R} dy' h_i h_i \right].
\ee

To extract angular momenta, we use
\bea
J_\phi &=& - \frac{\pi}{8 G_N} \int_0^{2 \pi R} dy' \ r^2 \frac{\delta g_{t'\phi}}{\sin^2 \theta},\\
J_\psi &=& - \frac{\pi}{8 G_N} \int_0^{2 \pi R} dy' \ r^2 \frac{\delta g_{t'\psi}}{\cos^2 \theta}.
\eea
A simple calculation then gives,
\bea
J_\phi &=&  \frac{\pi \lambda Q}{8 G_N} \int_0^{2 \pi R} dy' \left( \gamma_2 - \frac{\gamma_1 + \gamma_2}{2} \eta \left( 1- \frac{1}{4}h_i h_i - \frac{1}{\lambda^2}\right)\right)\\
&=&\frac{\pi \lambda Q}{8 G_N}  \gamma_2 (2 \pi R)~=~ \frac{n_1 n_5}{2} \left( m + \frac{1}{k}\right),
\eea
where we have used expression for $\lambda^{-2}$ \eqref{lambda} in going from the first to the second step. Similarly, we have
\bea
J_\psi ~=~\frac{\pi \lambda Q}{8 G_N}  \gamma_1 (2 \pi R)~=~ - \frac{n_1 n_5}{2}  m.
\eea

To summarise, the deformed state saturates the BPS bound and has charges
\begin{align}
& P_{y'} = \frac{n_1 n_5}{R} \left[ m \left( m + \frac{1}{k} \right)+ \frac{Q}{4a^2} \frac{1}{2\pi R} \int_0^{2\pi R} dy' h_i h_i \right], & 
J_\phi&= \frac{n_1 n_5}{2} \left( m + \frac{1}{k}\right),  \label{charges1} \\
& P_i = 0, &
J_\psi &= - \frac{n_1 n_5}{2}  m. \label{charges2}
\end{align}

\subsection{Smoothness}
Remarkably, the determinant of metric of the deformed solution gets no contribution from the scalars $\Phi_i$:
\be
\det g = - \frac{1}{4} \cos^2 \theta \sin^2 \theta h^2 f^2.
\ee
Therefore, as long as $\Phi_i$ remain finite, the potential singularities can only occur at places where the background geometry can become singular. The vicinity of these potentially dangerous points is analysed in \cite{gms2} for the undeformed solution. The analysis of that reference applies almost verbatim to our case together with the fact that the scalars \eqref{scalar_m} remain finite everywhere. 

This is perfectly in line with a conjecture of  reference \cite{Lunin:2012gp}.  They conjecture that any regular solution of the D1-D5 system can be deformed into a regular solution via the above technique provided, (i) $\Phi_i$ satisfies $\Box \Phi_i = 0$, (ii) $\Phi_i$ remains finite everywhere, (iii) $\Phi_i$ approaches a regular function $f_i(v)$ as $r \to  \infty$ on the four-dimensional base space.  Clearly all these conditions are met for the specific class of solutions studied in this paper.

\section{Identifying CFT states}
\label{sec:CFT}

\subsection{Decoupling limit}
To map the deformed geometries into states in the dual CFT, we need to evaluate charges in the AdS region rather than the asymptotically flat region. Such a computation is possible only when the deformed geometry has a large AdS region; and a decoupling limit can be taken. The geometry develops a large AdS region when we take 
\be
\epsilon \equiv \frac{a^2}{Q} \ll 1 \label{epsilon}.
\ee
To take the decoupling limit we must take $\e \to 0$ while keeping the AdS radius $\sqrt{Q}$ fixed. The relation \eqref{a_eq} implies that the size of the $y$-circle $R_y$ should go to infinity. We introduce 
\begin{align}
\bar u &= \frac{u}{R_y}, &
\bar v &= \frac{v}{R_y}, &
\bar r &= \frac{r}{a},
\end{align}
and take the limit $R_y \to \infty$. Without the deformation (i.e., with $\Phi_i = 0$) the decoupling limit gives
\bea 
ds^2&=&Q\left[-{\bar r}^2d{\bar u}d{\bar v}-
\frac{1}{4}(d{\bar u}+d{\bar v})^2+
\frac{d{\bar r}^2}{{\bar r}^2+k^{-2}}\right] \nn \\
&&+ \ Q\left[d\theta^2+
c_\theta^2\left(d\psi-\frac{1}{2k}(d{\bar u}-
d{\bar v})+md{\bar v}\right)^2+
s_\theta^2\left(d\phi-\frac{1}{2k}(d{\bar u}+
d{\bar v})-md{\bar v}\right)^2\right] \nn \\
& & + \ dz_i dz_i \,. \label{decoupled} 
\eea

To understand the decoupling limit with the scalars $\Phi_i$ turned on, we start by noting that in order to maintain ADM momentum \eqref{charges1} finite at $R_y \to \infty$,  we must scale the scalars $\Phi_i$ as 
\be
\Phi_i = \frac{a}{\sqrt{Q}} \,\bar \Phi_i = \frac{\sqrt{Q}}{R_y} \,\bar \Phi_i . 
\ee 
Then, in the metric, terms of the form 
\be
[du + \Phi_i dz_i]
\ee 
behave as 
\be 
du + \Phi_i dz_i
 = R_y \, d \bar u + \frac{\sqrt{Q}}{R_y} \, \bar \Phi_i \, dz_i ,
 \ee
 which in the decoupling limit  $R_y \to \infty$ simply becomes
 \be
  R_y \, d \bar u.
\ee
Thus, in effect, in the decoupling limit all $\Phi_i$ terms scale out, and we once again we get the decoupled metric \eqref{decoupled}. 

However, there is one subtlety. As we saw in the previous section the deformed metric is not manifestly asymptotically flat in coordinates $z_i, t, y$. It is better to change coordinates to  $z'_i, t', y'$ to connect the decoupled region to the asymptotically flat region. Through this change of coordinates the scalars reappear.  In order to implement these coordinate transformations,
we first observe that in the decoupling limit $\lambda$ from equation \eqref{lambda2} simplifies to unity,
\bea
\lambda^{-2} &=& \lim_{R_y \to \infty} \left[ 1 - \frac{1}{4}   \frac{Q}{R_y^2}  \left( \frac{1}{2 \pi R_y} \int_0^{2 \pi R_y} \bar f_i( \tilde v) \bar f_i( \tilde v) d \tilde v \right)\right] \nn \\
&=& 1.
\eea
Since $\lambda$ scales to unity, the transformations \eqref{diff1}--\eqref{diff3} simplify to 
\begin{align}
z'_i &= z_i - \frac{1}{2} \sqrt{Q}  \int_0^{\bar v} \bar \Phi_i \, d \bar{ \tilde{v}}, & u' &= u ,&
v' &=v.
\end{align} 

As a result, in primed coordinates the decoupled metric is 
\bea
ds^2&=&Q\left[-{\bar r}^2d{\bar u}d{\bar v}-
\frac{1}{4}(d{\bar u}+d{\bar v})^2+
\frac{d{\bar r}^2}{{\bar r}^2+k^{-2}}\right] \nn \\
&&+ \ Q\left[d\theta^2+
c_\theta^2\left(d\psi-\frac{1}{2k}(d{\bar u}-
d{\bar v})+md{\bar v}\right)^2+
s_\theta^2\left(d\phi-\frac{1}{2k}(d{\bar u}+
d{\bar v})-md{\bar v}\right)^2\right] \nn \\
& & + \ \left(dz'_i + \frac{1}{2}  \sqrt{Q}  \bar \Phi_i d \bar v \right)^2 \,. \label{decoupled2} 
\eea
We can now read off the charges. We find
\begin{align}
& P_{y'} = \frac{n_1 n_5}{R} \left[ m \left( m + \frac{1}{k} \right)+ \frac{1}{8\pi} \int_0^{2\pi} d\bar y \bar f_i \bar f_i \right], & 
J_\phi&= \frac{n_1 n_5}{2} \left( m + \frac{1}{k}\right),  \label{chargesAdS1} \\
& P_i = 0, &
J_\psi &= - \frac{n_1 n_5}{2}  m. \label{chargesAdS2}
\end{align}
These charges agree with \eqref{charges1}--\eqref{charges2} in the $R_y \to \infty$ limit.

\subsection{Deformed  states in the D1-D5 CFT}
The expression for the momentum $P_{y'}$, cf.~\eqref{chargesAdS1}, can be compared with momentum of the CFT state, 
\be
| \Psi \rangle = N \exp\left[ \sum_{n>0} \mu_n^i J_{-n}^i\right] |\psi \rangle,
\ee
where $|\psi \rangle$ is the undeformed state and $J_{-n}^i$ are the modes of the four U(1) currents of the D1-D5 CFT.
Assuming that the state   $|\psi \rangle$ is unit normalised, $\langle \psi | \psi \rangle = 1$,  we can fix the normalisation constant $N$ using the commutation relations,
\be
[J_{m}^i,J_{n}^j] = m \frac{n_1 n_5}{2} \delta^{ij} \delta_{m+n}.
\ee

Define $A^\dagger=  \sum_{n>0}  \mu_n^i J_{-n}^i$. Using  the fact  that the commutator
\be
[A, A^\dagger] = \frac{n_1 n_5}{2} \sum_{n>0} n  (\mu^i_n)^* \mu^i_n
\ee
 is a c-number, a small calculation shows that the normalisation constant $N$ is given by
 \be
1 = \langle \Psi | \Psi \rangle = N^2 \langle \psi | e^A  e^{A^\dagger} | \psi \rangle \\
= N^2 e^{[A, A^{\dagger}]}  \langle \psi |e^{A^\dagger}e^A| \psi \rangle =N^2 e^{[A, A^{\dagger}]},
\ee
where we have used $e^A |\psi \rangle = |\psi \rangle$ (which follows from $J_n^i |\psi\rangle =0$ for positive $n$).
This gives 
\be
N = \exp \left[ - \frac{n_1 n_5}{4} \sum_{n>0} n  (\mu^i_n)^* \mu^i_n\right].
\ee

To find the momentum, we compute the expectation value of $L_0$ and $\bar L_0$. Since right moving sector is untouched, we simply have 
\be
\langle \Psi | \bar L_0 | \Psi \rangle = \langle \psi | \bar L_0 | \psi \rangle. 
\ee
For the left sector, we need to do a computation. A simple way to organise this computation is as follows. Using the commutation relations,
\be
[L_m,J^i_n] = - n J^i_{m+n},
\ee
in particular, $[L_0,J^i_{-n}] =  n J^i_{-n}$, we get 
\be
[L_0 , A^\dagger] = \sum_{n>0} \mu_n^i [L_0, J_{-n}^i ]  = \sum_{n>0} n \mu_n^i  J^i_{-n} =: B^\dagger.
\ee
To calculate $\langle \Psi |  L_0 | \Psi \rangle$ we observe
\be
\langle \Psi |  L_0 | \Psi \rangle = N^2 \langle \psi | e^A L_0 e^{A^\dagger} | \psi \rangle 
= N^2  \langle  \psi | e^A  e^{A^\dagger}e^{-A^\dagger}L_0e^{A^\dagger}|\psi \rangle.  
\ee
Now we can use Baker–Campbell–Hausdorff formula to write $e^{-A^\dagger}L_0 e^{A^\dagger} = L_0 + B^\dagger$. We also use $e^A e^{A^\dagger}= e^{A^\dagger}e^{A} e^{[A,A^\dagger]}$ and the fact that $N^2 e^{[A,A^\dagger]}=1$ as shown earlier. We get 
\be
\langle \Psi |  L_0 | \Psi \rangle = N^2  \langle  \psi | e^A  e^{A^\dagger}(L_0 + B^\dagger)|\psi \rangle 
=   \langle  \psi | e^{A^\dagger}e^A (L_0 + B^\dagger)|\psi \rangle.
\ee
Now we use $[L_0,A] |\psi\rangle = B|\psi \rangle =0$, as $B$ contains only $J^{i}_{n}$ with positive $n$, we get 
\bea
\langle \Psi |  L_0 | \Psi \rangle &=&  \langle \psi | L_0 | \psi \rangle +   \langle \psi | [A, B^\dagger] | \psi \rangle \\
 &=& \langle \psi | L_0 | \psi \rangle +   \sum_{ n> 0} \frac{n^2 n_1 n_5}{2} (\mu^i_n)^* \mu^i_n,
\eea
We conclude that,
\bea
\langle \Psi |  L_0 - \bar L_0 | \Psi \rangle  = R P_{y'} =  \langle \psi | L_0 - \bar L_0 | \psi \rangle +  \sum_{ n> 0} \frac{n^2 n_1 n_5}{2}( \mu^i_n)^* \mu^i_n.
\eea
Upon doing the Fourier expansion of \eqref{chargesAdS1} in the decoupling limit, we get
\be
R P_{y'} = \langle \psi | L_0 - \bar L_0 | \psi \rangle +   \sum_{ n> 0} \frac{n_1 n_5}{2} \frac{Q}{a^2}\left(  (c^i_n)^* c^i_n \right).
\ee
Therefore, the map between the quantities $c_n^i$ and $\mu_n^i$ is
\be
\mu_n^i  = \frac{1}{n} \sqrt{\frac{Q}{a^2}}c_n^i. 
\ee

Let us remark that in the computations of this subsection the only property of the undeformed state $|\psi\rangle$ we have used is that it is annihilated by $A$ and $B$ operators. The above analysis is therefore applicable to a  large class of states. Although matching of the charges is no proof that the identified states are dual to the gravity deformation considered above; it is a strong indicator.

\section{Dualities and the generalized Garfinkle-Vachaspati transform}
\label{sec:dualities}
In an attempt to explore further applications of the generalized Garfinkle-Vachaspati transform and related solution generating techniques, in this section we write deformed Bena-Warner solutions in various M2-M5-P duality frames. We obtain these various duality frames by applying dualities. Our starting point is the D1-D5-P frame. In appendix \ref{sec:dualities_details} the dictionary for going from the M2-M2-M2 BW form to the D1-D5-P form is worked out. The string frame D1-D5-P metric can be written in the following form, cf.~\eqref{metric_after_T_dualities},
\be
ds_{10}^2=-\frac{1}{Z_3 Z_1}(dt+\k)^2+Z_1 h_{mn}dx^m dx^n  +\frac{Z_3}{Z_1}(dz_5+A^{(3)}_\mu dx^\mu )^2+ (dz_1^2 +dz_2^2+ dz_3^2 + d z_4^2), \label{metric_after_T_dualities:main_text}
\ee
where
\be
A^{(3)}_\mu dx^\mu = -\frac{dt+\k}{Z_3}+\omega_3.
\ee
The RR two-form field supporting this solution takes the form, cf.~\eqref{2formpotential},
\be
C=-\left(\frac{dt+\k}{Z_1}-\omega_1\right)\wedge (dz_5+\omega_3)+ \sigma \label{2formpotential:main_text}.
\ee
where the two-form $\sigma$ satisfies equation \eqref{sigma_eq}.

The application of the generalized Garfinkle-Vachaspati transform  with,
\begin{align}
&k = \frac{\partial}{\partial t}, & & l = \frac{\partial}{\partial z_4}, \label{GV:vecs} \\
&k_\mu dx^\mu = -Z_1^{-1}(dz_5+\omega_3), & & l_\mu dx^\mu = d z_4,
\end{align}
leads to the transformed metric,
\begin{eqnarray}
\nonumber
(ds_{10}')^2= ds^2_{10} - 2 Z_1^{-1}\Phi (dz_5+\omega_3)dz_4,
\end{eqnarray}
with the transformed $C$-field,
\be
C' =C+\frac{\Phi}{Z_1} (dz_5+\omega_3) \wedge dz_4.
\ee
These deformed solutions we now write in various other duality frames.

\subsection*{T-duality along $z_1$-direction and M-theory lift}

The first duality frame we explore is obtained by T-duality along $z_1$-direction followed by an M-theory lift along $z_6$:
\be
\nn
\mathrm{D1}_{z_5}-\mathrm{D5}_{z_1 z_2 z_3 z_4 z_5}-\mathrm{P}_{z_5}\xrightarrow{\mathrm{T}_{z_1}} 
\mathrm{D2}_{z_1 z_5}-\mathrm{D4}_{z_2 z_3 z_4 z_5}-\mathrm{P}_{z_5}\xrightarrow{\text{M-theory lift}} 
\mathrm{M2}_{z_1 z_5}-\mathrm{M5}_{z_2 z_3 z_4 z_5 z_6}-\mathrm{P}_{z_5}.
\ee
Performing these dualities, the final answer for the metric is
\begin{eqnarray}
\nonumber
ds_{11}^2=ds^2_{10} - 2 Z_1^{-1}\Phi (dz_5+\omega_3)dz_4 + dz_6^2,
\end{eqnarray}
together with the 3-form field
\be
\mathcal{A}^{(3)}=
\left(C+\frac{\Phi}{Z_1} (dz_5+\omega_3) \wedge dz_4\right) \wedge dz_1.
\ee
In this duality frame, the transformation is essentially of the form of the generalised Garfinkle-Vachaspati transform. It is natural to conjecture that a solution generating technique akin to generalised Garfinkle-Vachaspati transform exist in (an appropriate truncation of) M-theory.

\subsection*{T-dualities along $z_1,z_2,z_3$ and M theory lift}
The next duality frame we explore is obtained by T-dualities along $z_1,z_2,z_3$-directions followed by an M-theory lift along $z_6$:
\be
\nn
\mathrm{D1}_{z_5}-\mathrm{D5}_{z_1 z_2 z_3 z_4 z_5}-\mathrm{P}_{z_5}\xrightarrow{\mathrm{T}_{z_1 z_2 z_3}} 
\mathrm{D4}_{z_1 z_2 z_3 z_5}-\mathrm{D2}_{z_4 z_5}-\mathrm{P}_{z_5}\xrightarrow{\text{M-theory lift}} 
\mathrm{M5}_{z_1 z_2 z_3 z_5 z_6}-\mathrm{M2}_{z_4 z_5}-\mathrm{P}_{z_5}.
\ee
Performing these dualities, the eleven-dimensional metric is,
\begin{eqnarray}
\nonumber
ds_{11}^2=ds^2_{10} - 2 Z_1^{-1}\Phi (dz_5+\omega_3)dz_4 + dz_6^2,
\end{eqnarray}
together with the $\mathcal{A}^{(6)}$ in eleven-dimensions, which is thought of as the dual of $\mathcal{A}^{(3)}$: 
\be
\mathcal{A}^{(6)}=
\left(C+\frac{\Phi}{Z_1} (dz_5+\omega_3) \wedge dz_4\right)  \wedge dz_1 \wedge dz_2 \wedge dz_3 \wedge dz_6.
\ee
Even in this duality frame, the transformation is essentially of the form of the generalised Garfinkle-Vachaspati transform. Once again, it is natural to conjecture that a solution generating technique akin to generalised Garfinkle-Vachaspati transform exist in such a set-up.

\subsection*{T-duality along $z_4$-direction and M-theory lift}

The next duality frame we explore is obtained by T-duality along $z_4$-directions followed by an M-theory lift along $z_6$. Recall that $z_4$ is also the spacelike direction used for the generalised Garfinkle-Vachaspati transform, cf.~\eqref{GV:vecs}. The duality sequence is:
\be
\nn
\mathrm{D1}_{z_5}-\mathrm{D5}_{z_1 z_2 z_3 z_4 z_5}-\mathrm{P}_{z_5}\xrightarrow{\mathrm{T}_{z_4}} 
\mathrm{D2}_{z_4 z_5}-\mathrm{D4}_{z_1 z_2 z_3 z_5}-\mathrm{P}_{z_5}\xrightarrow{\text{M-theory lift}} 
\mathrm{M2}_{z_4 z_5}-\mathrm{M5}_{z_1 z_2 z_3 z_5 z_6}-\mathrm{P}_{z_5}.
\ee
After the T-duality the IIA  ten-dimensional metric in the string frame is,
\be
ds_{10}^2 = -2Z_1^{-1}(dt+k)(dz_5 + \omega_3) + \frac{Z_3}{Z_1}\left(1-\frac{\Phi^2}{Z_1Z_3}\right) (dz_5+\omega_3)^2 + Z_1 h_{mn}dx^m dx^n + ds^2_{\mathrm{T}^4}.
\ee
The associated form-fields are,
\begin{align}
&C^{(3)}=C \wedge dz_4,& & C^{(1)}=\frac{\Phi}{Z_1}(dz_5+\omega_3),&
&B^{(2)}=\frac{\Phi}{Z_1}(dz_5+\omega_3)\wedge dz_4.
\end{align}
The dilaton remains the same, i.e., $e^{2\phi}=1$. The M-theory lift is,
\bea
ds_{11}^2 &=& ds^2_{10} + \frac{2\Phi}{Z_1}(dz_5 + \omega_3) dz_6 + dz_6^2, 
\\ 
\mathcal{A}^{(3)} &=& C^{(3)} + \frac{\Phi}{Z_1}(dz_5+\omega_3)\wedge dz_4 \wedge dz_6 .
\eea
In this duality frame too, the transformation is essentially of the generalised Garfinkle-Vachaspati form.

Similarly, one can consider another 
 duality chain to another M2-M5-P frame as follows
\be
\nn
\mathrm{D1}_{z_5}-\mathrm{D5}_{z_1 z_2 z_3 z_4 z_5}-\mathrm{P}_{z_5}\xrightarrow{\mathrm{T}_{z_1 z_2 z_4}} 
\mathrm{D4}_{z_1 z_2 z_4 z_5}-\mathrm{D2}_{z_3 z_5}-\mathrm{P}_{z_5}\xrightarrow{\text{M-theory lift}} 
\mathrm{M5}_{z_1 z_2 z_4 z_5 z_6}-\mathrm{M2}_{z_3 z_5}-\mathrm{P}_{z_5}.
\ee
Even in this duality frame the transformation is essentially of the
Garfinkle-Vachaspati form. It is tempting to speculate that some  
solution generating techniques akin to generalised Garfinkle-Vachaspati transform exist for these set-ups as well.

\section{Conclusions and future directions}
\label{sec:disc}

In this paper, we have presented generalized Garfinkle-Vachaspati transform as a solution generating 
technique and have  analysed in detail corresponding deformations of certain D1-D5-P orbifolds.  We considered states that are obtained by (odd) integeral spectral flows on certain NS sector chiral primaries.  A more general supersymmetric family is known where the spectral flow parameter is also fractionated \cite{Jejjala:2005yu, Giusto:2012yz, Chakrabarty:2015foa}. We expect our deformation analysis to straightforwardly  extend to that setting as well. A much more difficult question is how to add a similar deformation to non-supersymmetric solutions considered in \cite{Jejjala:2005yu, Chakrabarty:2015foa}.  The analysis of the current paper does not seem to be applicable, since in general such solutions do not admit null Killing vector. It will be  interesting to figure out if a variant of the above analysis can be applied.\footnote{A different, but related, type of deformation on the simplest of non-supersymmetric solutions of \cite{Jejjala:2005yu} was studied in \cite{Roy:2016zzv}. It is tempting to speculate, given the analysis \cite{Mathur:2011gz, Roy:2016zzv}, that a variant of the above analysis finds application to non-supersymmetric settings.}

In the paper, we only considered deformation of solutions of minimal six-dimensional supergravity embedded in ten-dimensional IIB theory. Extension to non-minimal six-dimensional supergravity in a natural direction to explore.  A form of such deformation for supersymmetric solutions was proposed in \cite{Lunin:2012gp}. It will be interesting to check the validity of the proposed form and to relate it  to our generalised Garfinkle-Vachaspati transform. 

In an attempt to explore further applications of the generalized Garfinkle-Vachaspati transform, in section \ref{sec:dualities} we wrote a class of deformed solutions in various M2-M5-P duality frames.   It is natural to speculate that some  
variant of the generalised Garfinkle-Vachaspati transform also exist for these M-theory set-ups.

Our generalized Garfinkle-Vachaspati transformation is an example of the extended Kerr-Schild metrics considered in \cite{Ett:2010by} and \cite{Malek:2014dta}. Due to the assumption that the null and spacelike vectors are Killing, our analysis is more restrictive and hence our final results are much simpler.  In addition, we have non-trivial matter  present compared to the general extended Kerr-Schild forms considered in those references. It will be interesting to see if we can further relax our conditions on null and spacelike vectors and relate our analysis to theirs. 
 
Since the number of Killing symmetries do not change under our generalized Garfinkle-Vachas-pati deformation, it is natural to ask whether the deformation has a simple group theory interpretation from the hidden symmetry point of view of type IIB theory. Hidden symmetries under null reduction of gravity theories have not been fully explored. Some general results are known \cite{Julia:1994bs}. It can be useful to explore the null reduction further and find the interpretation of (generalised) Garfinkle-Vachaspati transform from the hidden symmetry point of view.  We hope to return to some of the above problems in our future work.

\subsection*{Acknowledgements} We thank Swayamsidha Mishra, Ashoke Sen, David Turton, and especially Oleg Lunin for discussions. AV thanks NISER Bhubaneswar, AEI Potsdam, and ICTP Trieste for warm hospitality towards the final stages of this project. The work of AV is supported in part  by the DST-Max Planck Partner Group ``Quantum Black Holes'' between CMI Chennai and AEI Potsdam. 

\appendix

\section{Detailed analysis of the equations of motion}
\label{sec:GV_details}
We establish that generalised Garfinkle-Vachaspati transform is a valid solution generating technique via a brute force calculation. We show that the left and the right hand side of the Einstein equations transform in the exactly the same way, thereby establishing that if we start with a solution, we can deform it to a new solution. In our convention, Einstein equations are 
\begin{equation}
R_{\mu\nu} =\frac{1}{4}F_{\mu\lambda\sigma}{F_\nu}^{\lambda\sigma} \label{einstein_eqs},
\end{equation}
and 
matter field equations are 
\be
\del_\mu F^{\mu \nu \rho} =0. 
\ee
The tedious calculations required to show that these equations transform covariantly are organised as follows: in section \ref{sec:left} the left hand side of the Einstein equations are analysed, 
in section \ref{sec:right} the right hand side of the Einstein equations are analysed, and finally
in section \ref{sec:matter} matter equations are analysed. 

The generalised Garfinkle-Vachaspati transform of the  metric is,
\be
g'_{\mu\nu}=g_{\mu\nu}+\Phi (k_{\mu}l_{\nu}+k_\nu l_\mu), \label{transformed_metric}
\ee
where $\Phi$ is a massless scalar on the original background spacetime $g_{\mu\nu}$,
\be
\Box\Phi=0. \label{massless_scalar}
\ee
The vector $k^\mu$ appearing in  \eqref{transformed_metric} is a null Killing vector
\begin{align}
k^\mu k_\mu &= 0, &
\nabla_\mu k_\nu+\nabla_\nu k_\mu &=0,
\end{align}
and $l^\mu$ is a unit normalised covariantly constant spacelike (Killing) vector orthogonal to $k^\mu$:
\begin{align}
l^\mu l_\mu &=1, &
k^\mu l_\mu &=0, &
\nabla_\mu l_\nu&=0.
\end{align}
Furthermore, we also require that the scalar $\Phi$ is compatible with the Killing symmetries, 
\begin{align}
k^\mu \nabla_\mu \Phi &=0, &
l^\mu \nabla_\mu \Phi &=0,
\end{align}
so that the transformed spacetime $g'_{\mu \nu}$ also has $k^\mu$ and $l^\mu$ as  Killing symmetries.

\subsection{Left hand side of Einstein equations}
\label{sec:left}
The aim of this subsection is to find the transformation of the left hand side of the Einstein equations \eqref{einstein_eqs}. Doing this  is straightforward, though somewhat tedious. To compute the change in the Ricci tensor, we essentially need to compute the change in the metric compatible connection and its covariant derivative:
\be
 R'_{\lambda\nu}=R_{\lambda\nu}-{\nabla}_\lambda {{\Omega}^{\mu}}_{\mu\nu} + {\nabla}_{\mu} {{\Omega}^{\mu}}_{\lambda\nu} +{\Omega^\mu}_{\mu\rho}
{\Omega^\rho}_{\lambda\nu} -{\Omega^\rho}_{\mu\lambda}{\Omega^\mu}_{\rho\nu},
\label{ricci_transformed}
\ee
where $\Omega^\mu _{\lambda\nu}$ is the change in the metric compatible connection
\be
\Gamma{'}^\mu _{\lambda\nu}=\Gamma^\mu _{\lambda\nu}+\Omega^\mu _{\lambda\nu}.
\ee
The change in the metric compatible connection is 
\be
\Omega^\mu _{\lambda\nu} = \frac{1}{2}g'^{\mu \sigma}\left(\del_\lambda g'_{\nu \sigma} + \del_\nu g'_{\sigma \lambda }- \del_\sigma g'_{\nu \lambda }\right).
\ee
We compute various pieces required in equation \eqref{ricci_transformed}.

We start by observing that the 
inverse of the transformed metric \eqref{transformed_metric} is simply
\be
g'^{\mu\nu}=g^{\mu\nu}+\Phi^2 k^\mu k^\nu-\Phi S^{\mu\nu}.
\ee
Next, we introduce the notation,
\begin{eqnarray}
S_{\mu\nu} &=& k_\mu l_\nu + k_\nu l_\mu, \\
h_{\mu\nu} &=& \Phi S_{\mu\nu},\\
n_{\mu\nu} &=& \del_\mu k_\nu -\del_\nu k_\mu.
\end{eqnarray}
The change in the metric compatible connection, $\Omega^\mu _{\lambda\nu}$, is conveniently organised in two terms,
\be
\Omega^\mu _{\lambda\nu}={\Xi}^\mu _{\lambda\nu}+\frac{1}{2}(\Phi^2 k^\mu k^\alpha-\Phi S^{\mu\alpha})(\del_\lambda h_{\nu\alpha}+\del_\nu h_{\alpha\lambda}-\del_\alpha h_{\lambda\nu}),
\ee
where the  first term ${\Xi}^\mu _{\lambda\nu}$ is the combination that features in the Garfinkle-Vachaspati transform without the spacelike Killing vector $l^\mu$ \cite{Kaloper:1996hr}: 
\be
{\Xi}^\mu _{\lambda\nu}=\frac{1}{2}g^{\mu\alpha}(\del_\lambda h_{\nu\alpha}+\del_\nu h_{\alpha\lambda}-\del_\alpha h_{\lambda\nu}).
\ee
In order to proceed further we make a convenient definition,
\be 
K^\mu_{\nu\lambda}:=\del_\nu S^\mu_\lambda+\del_\lambda S^\mu_\nu-\del^\mu S_{\lambda\nu},
\ee
using which it follows that
\be 
{\Xi}^\mu _{\lambda\nu}=
\frac{1}{2}\left((\del_\nu\Phi) S^\mu_\lambda+(\del_\lambda\Phi) S^\mu_\nu-(\del^\mu\Phi) S_{\nu\lambda}+\Phi K^\mu_{\nu\lambda}\right),
\ee
and therefore,
\be
\Omega^\mu _{\lambda\nu}={\Xi}^\mu _{\lambda\nu}-\frac{1}{2}\Phi k^\mu (k_\nu \del_\lambda\Phi+k_\lambda\del_\nu\Phi).
\label{Omega_Xi}
\ee
The trace of $ \Omega^\mu _{\lambda\nu}$is easily seen to be zero
\be
\Omega^\mu _{\mu\lambda}=0.
\ee
As a result the transformation of the Ricci tensor \eqref{ricci_transformed} simplifies to 
\be
 R'_{\lambda\nu}=R_{\lambda\nu}+ {\nabla}_{\mu} {{\Omega}^{\mu}}_{\lambda\nu}  -{\Omega^\rho}_{\mu\lambda}{\Omega^\mu}_{\rho\nu}. \label{simp_ricci}
\ee
To compute the right hand side of the above expression, we need to compute 
${\nabla}_{\mu} {{\Omega}^{\mu}}_{\lambda\nu}$ and ${\Omega^\rho}_{\mu\lambda}{\Omega^\mu}_{\rho\nu}$. 
We can first show that
\bea
2\del_\mu {\Xi}^\mu _{\lambda\nu}=(\del_\mu \del_\nu\Phi)S^\mu_\lambda+(\del_\mu \del_\lambda\Phi)S^\mu_\nu
-(\del^\mu\Phi)(\del_\mu S_{\nu\lambda})+(\del_\mu\Phi) K^\mu_{\nu\lambda}
+\Phi (\del_\mu K^\mu_{\nu\lambda}) \label{del_Xi},
\eea
where we have used $\del_\mu S^\mu_\nu=0$ and the fact that we are deforming the original solution via a massless scalar field \eqref{massless_scalar}.
The first three terms of \eqref{del_Xi} combine to zero,
\bea
(\del_\mu\del_\nu\Phi)S^\mu_\lambda+(\del_\mu \del_\lambda\Phi)S^\mu_\nu-(\del^\mu\Phi)(\del_\mu S_{\nu\lambda})=0.
\eea
In order to simplify \eqref{del_Xi} further we develop some identities. One can easily show that
\bea
K^\mu_{\nu\lambda}&=&(\del_\nu k^\mu-\del^\mu k_\nu)l_\lambda+(\del_\lambda k^\mu -\del^\mu k_\lambda) l_\nu\\
&=&{n_\nu}^\mu l_\lambda + {n_\lambda}^\mu l_\nu. \label{K_simp}
\eea
It then follows that the fourth term of \eqref{del_Xi} simplifies to 
\bea
(\del_\mu\Phi)K^\mu_{\nu\lambda}&=&-2 k^\mu [(\del_\nu\del_\mu \Phi)l_\lambda+(\del_\lambda\del_\mu \Phi)l_\nu ],
\eea
where we have used
\bea
(\del_\mu\Phi){n_\nu}^\mu &=&-2 k^\mu (\del_\nu\del_\mu \Phi).
\eea
Inserting \eqref{K_simp} in  $(\del_\mu K^\mu_{\nu\lambda})$, the last term of \eqref{del_Xi} simplifies to
\bea
\del_\mu  K^\mu_{\nu\lambda}=-2(\square k_\nu)l_\lambda-2(\square k_\lambda)l_\nu,
\eea
where we have also used 
\bea
\del_\mu {n_\nu}^\mu =-2\square k_\nu.
\eea
When the dust settles, we get a simplified expression for equation \eqref{del_Xi}:
\be
\del_\mu {\Xi}^\mu _{\lambda\nu}=-l_\lambda[ k^\mu (\del_\nu\del_\mu \Phi)+\Phi \square k_\nu ]-l_\nu [ k^\mu (\del_\lambda\del_\mu \Phi)+\Phi \square k_\lambda ] .
\label{del_X_simp}
\ee
From \eqref{Omega_Xi} it then follows that
\be
2\del_\mu \Omega^\mu _{\lambda\nu}=2\del_\mu {\Xi}^\mu _{\lambda\nu}-\Phi k^\mu [k_\nu (\del_\mu\del_\lambda\Phi)+k_\lambda (\del_\mu\del_\nu\Phi)].
\ee

This is one of the pieces that is required to compute the change in the Ricci tensor \eqref{simp_ricci}. The other piece that is required is $\Omega^\rho_{\mu\lambda}\Omega^\mu_{\rho\nu}$. In order to compute this combination, we start by observing that
\bea
4\Omega^\rho_{\mu\lambda}\Omega^\mu_{\rho\nu}&=&[2{\Xi}^\rho_{\mu\lambda}-\Phi k^\rho (k_\mu\del_\lambda\Phi+k_\lambda\del_\mu\Phi)][2{\Xi}^\mu_{\rho\nu}-\Phi k^\mu (k_\rho\del_\nu\Phi+k_\nu\del_\rho\Phi)]\\
&=&4{\Xi}^\rho_{\mu\lambda}{\Xi}^\mu_{\rho\nu}.
\eea
The combination ${\Xi}^\rho_{\mu\lambda}{\Xi}^\mu_{\rho\nu}$ is,
\bea
4{\Xi}^\rho_{\mu\lambda}{\Xi}^\mu_{\rho\nu}&=&[(\del_\mu\Phi)S^\rho_\lambda+(\del_\lambda\Phi)S^\rho_\mu-(\del^\rho\Phi)S_{\mu\lambda}+\Phi K^\rho_{\mu\lambda}]
\nonumber
\\ && \quad \times \ [(\del_\rho\Phi)S^\mu_\nu+(\del_\nu\Phi)S^\mu_\rho-(\del^\mu\Phi)S_{\rho\nu}+\Phi K^\mu_{\rho\nu}.
\eea
In order to simplify this further, we use the following non-trivial identities, which can be straightforwardly established: 
\begin{align}
S^\rho_\mu K^\mu_{\rho\nu}&=0, &
S_{\mu\lambda}K^\mu_{\rho\nu}&=0,\\
S^{\mu}_{\nu}K^\rho_{\mu\lambda}&=k_\nu n_\lambda{}^\rho, &
K^\rho_{\mu\lambda}K^\mu_{\rho\nu}&=4(\del_\mu k^\rho)(\del_\rho k^\mu ) l_\lambda l_\nu.
\end{align}
After all these simplifications, we get 
\be
\Omega^\rho_{\mu\lambda}\Omega^\mu_{\rho\nu}=-\frac{1}{2}(\del_\rho\Phi)(\del^\rho\Phi)k_\lambda k_\nu-\frac{1}{2}\Phi k^\mu [k_\lambda (\del_\mu \del_\nu \Phi)+k_\nu (\del_\mu\del_\lambda\Phi)]+\Phi^2(\del_\mu k^\rho)(\del_\rho k^\mu ) l_\lambda l_\nu.
\ee
Therefore, a final simplified expression for the transformed Ricci tensor is
\bea
R'_{\lambda\nu} 
 &=& R_{\lambda\nu}-l_\lambda[ k^\mu (\del_\nu\del_\mu \Phi)+\Phi \square k_\nu ]-l_\nu [ k^\mu (\del_\lambda\del_\mu \Phi)+\Phi \square k_\lambda ]\nonumber \\
&&+\frac{1}{2}(\del_\rho\Phi)(\del^\rho\Phi)k_\lambda k_\nu-\Phi^2(\del_\mu k^\rho)(\del_\rho k^\mu ) l_\lambda l_\nu.
\eea
In the next subsection we show that the right hand side of the Einstein equations \eqref{einstein_eqs} also transform in the same way.

 \subsection{Right hand side of Einstein equations}
 \label{sec:right}
We start by
recalling that under generalised Garfinkle-Vachaspati transform the two-form field transforms as
\be
C \to C'=C-\Phi \ k_\mu dx^\mu \wedge l_\nu dx^\nu.
\ee
To show that the right hand side of the Einstein equations \eqref{einstein_eqs} transform in the same way, we require 
\begin{eqnarray}
k^\mu {F_\mu}^{\nu\rho}=-n^{\nu\rho}, \label{condition_C}
\end{eqnarray}
and
\begin{eqnarray}
l^\mu {F_\mu}^{\nu\rho}=0. \label{condition_C_l}
\end{eqnarray}

As mentioned in the main text, these conditions are satisfied by a large class of solutions of the minimal six-dimensional supergravity embedded in type IIB theory.
Introducing the notation
 \bea
 m_{\mu\nu} &=& k_\mu l_\nu - k_\nu l_\mu, \label{notation_m}
 \eea
we have
\bea
C'_{\mu \nu } &=&C_{\mu \nu}-\Phi(k_\mu l_\nu-k_\nu l_\mu) \\ &=&C_{\mu \nu}-\Phi m_{\mu \nu}.
\eea
It then simply follows that
\bea
F'_{\mu\nu\rho}&=&\partial_\mu C_{\nu\rho}+\partial_\rho C_{\mu\nu}+\partial_\nu C_{\rho\mu }-\partial_\mu(\Phi m_{\nu\rho})-\partial_\rho(\Phi m_{\mu\nu})-\partial_\nu(\Phi m_{\rho\mu })\\
&=&\partial_\mu C_{\nu\rho}+\partial_\rho C_{\nu\mu }+\partial_\mu C_{\rho\nu}-Q_{\mu\nu\rho}-\Phi P_{\mu\nu\rho}\\
&=&F_{\mu\nu\rho}-Q_{\mu\nu\rho}-\Phi P_{\mu\nu\rho},
\end{eqnarray}
where
\begin{eqnarray}
Q_{\mu\nu\rho}&=&(\partial_\mu\Phi)m_{\nu\rho}+(\partial_\rho\Phi)m_{\mu\nu}+(\partial_\nu\Phi)m_{\rho\mu }, \label{Q}\\
P_{\mu\nu\rho}&=&\partial_\mu m_{\nu\rho}+\partial_\rho m_{\mu\nu}+\partial_\nu m_{\rho\mu }. \label{P}
\end{eqnarray}
Inserting \eqref{notation_m} in \eqref{P} we get,
\begin{eqnarray}
P_{\mu\nu\rho}&=&\partial_\mu (k_\nu l_\rho - k_\rho l_\nu)+\partial_\rho (k_\mu l_\nu - k_\nu l_\mu )+\partial_\nu (k_\rho l_\mu  - k_\mu l_\rho)\\
&=& (\partial_\mu k_\nu-\partial_\nu k_\mu )l_\rho+(\partial_\rho k_\mu -\partial_\mu k_\rho)l_\nu+(\partial_\nu k_\rho-\partial_\rho k_\nu)l_\mu \\
&=&n_{\mu\nu}l_\rho+n_{\rho\mu }l_\nu+n_{\nu\rho}l_\mu .
\end{eqnarray}

To compute the transformed right hand side of the Einstein equations, we need to first raise the indices on the three-form field $F_{\mu \nu \lambda}$. Raising the first index we get, 
\begin{eqnarray}
{F'^\sigma}_{\nu\rho}&=&g'^{\mu\sigma}F'_{\mu\nu\rho}\\
&=&(g^{\mu\sigma}+\Phi^2 k^\mu k^\sigma -\Phi S^{\mu\sigma})(F_{\mu\nu\rho}-Q_{\mu\nu\rho}-\Phi P_{\mu\nu\rho}).
\end{eqnarray}
Using the identities, 
\begin{eqnarray}
k^\mu Q_{\mu\nu\rho}&=&0,\\
k^\mu P_{\mu\nu\rho}&=&0,\\
S^{\mu\sigma} F_{\mu\nu\rho}&=&-l^\sigma n_{\nu\rho},\\
S^{\mu\sigma}Q_{\mu\nu\rho}&=&k^\sigma[k_\rho(\partial_\nu\Phi)-k_\nu(\partial_\rho\Phi)],\\
S^{\mu\sigma}P_{\mu\nu\rho}&=&k^\sigma n_{\nu\rho},
\end{eqnarray}
it follows that,
\begin{eqnarray}
{F'^\sigma}_{\nu\rho}&=&
{F^\sigma}_{\nu\rho}-{Q^\sigma}_{\nu\rho}-\Phi {P^\sigma}_{\nu\rho}+\Phi l^\sigma n_{\nu\rho}+\Phi k^\sigma [(\partial_\nu\Phi)k_\rho-(\partial_\rho\Phi)k_\nu].
\end{eqnarray}
Similarly raising the second index we get,
\begin{eqnarray}
\nonumber
{F'^{\sigma\eta}}_\rho&=&{g'}^{\eta\nu}{F'^\sigma}_{\nu\rho}\\
&=&{F^{\sigma\eta}}_\rho-{Q^{\sigma\eta}}_\rho-\Phi {P^{\sigma\eta}}_\rho+\Phi l^\sigma {n^\eta}_\rho+\Phi k^\sigma [(\partial^\eta\Phi)k_\rho-(\partial_\rho\Phi)k^\eta] \nn \\ 
&&-\Phi l^\eta ({n^\sigma}_\rho)-\Phi k^\eta [(\partial^\sigma\Phi)-k^\sigma(\partial_\rho\Phi)].
\end{eqnarray}

Given the above expressions, it is possible to compute the change in the right hand side of the Einsteins equations. However, it turns out that for various purposes the three-form with all three indices raised is a much easier quantity to work with. We now write an expression for $F'$ with all three indices raised, and then turn to Einstein equations. We have 
\begin{eqnarray}
F'^{\sigma\eta\alpha}&=&g'^{\rho\alpha}{F'^{\sigma\eta}}_\rho\\
\nonumber
&=&F^{\sigma\eta\alpha}-Q^{\sigma\eta\alpha}-\Phi P^{\sigma\eta\alpha}+\Phi l^\sigma n^{\eta\alpha}+\Phi k^\sigma[(\partial^\eta\Phi)k^\alpha-(\partial^\alpha\Phi)k^\eta]\\
\nonumber
&&-\Phi l^\eta (n^{\sigma\alpha})-\Phi k^\eta [k^\alpha(\partial^\sigma \Phi)-k^\sigma (\partial^\alpha \Phi)])+\Phi^2 k^\alpha k^\rho {F^{\sigma\eta}}_\rho-\Phi S^{\alpha\rho}{F^{\sigma\eta}}_\rho\\
&&+\Phi S^{\alpha\rho}{Q^{\sigma\eta}}_\rho+\Phi^2 S^{\alpha\rho}{P^{\sigma\eta}}_\rho\\
\nonumber &=&F^{\sigma\eta\alpha}-Q^{\sigma\eta\alpha}-\Phi (n^{\sigma\eta}l^\alpha+n^{\alpha\sigma}l^\eta+n^{\eta\alpha}l^\sigma)+\Phi l^\sigma n^{\eta\alpha}\\
\nonumber
&&+\Phi k^\sigma[(\partial^\eta\Phi)k^\alpha-(\partial^\alpha\Phi)k^\eta] -\Phi l^\eta (n^{\sigma\alpha})-\Phi k^\eta [k^\alpha(\partial^\sigma\Phi)-k^\sigma (\partial^\alpha\Phi)])\\
\nonumber
&&+\Phi l^\alpha (n^{\sigma\eta}) +\Phi k^\alpha [k^\eta(\partial^\sigma\Phi)-k^\sigma (\partial^\eta\Phi)]\\
\nonumber
&=&F^{\sigma\eta\alpha}-Q^{\sigma\eta\alpha}+\Phi k^\sigma [(\partial^\eta\Phi)k^\alpha-(\partial^\alpha\Phi)k^\eta]-\Phi k^\eta [k^\alpha(\partial^\sigma\Phi)-k^\sigma(\partial^\alpha\Phi)])\\
&&+\Phi k^\alpha [k^\eta(\partial^\sigma\Phi)-k^\sigma(\partial^\eta\Phi)]\\
&=&F^{\sigma\eta\alpha}-Q^{\sigma\eta\alpha},
\end{eqnarray}
which is a remarkably simple equation.

Now we are in position to compute the transformed right hand side of \eqref{einstein_eqs}. Using identities
\begin{eqnarray}
-F_{\lambda\alpha\beta}{Q}^{\delta\alpha\beta}-Q_{\lambda\alpha\beta}{F}^{\delta\alpha\beta}&=&-4[l^\delta (\del_\lambda\del_\beta\Phi) +l_\lambda(\del^\delta \del_\beta\Phi)]k^\beta,\\
Q_{\lambda\alpha\beta}{Q}^{\delta\alpha\beta}&=&2(\partial_\beta\Phi)(\partial^\beta\Phi)k_\lambda k^\delta, \\
P_{\lambda\alpha\beta}{Q}^{\delta\alpha\beta}&=&4 k^\delta k^\alpha (\del_\lambda\del_\alpha\Phi),\\
P_{\lambda\alpha\beta}F^{\delta\alpha\beta}&=&4l_\lambda \square k^\delta,
\end{eqnarray}
we get,
\bea
\frac{1}{4}F'_{\lambda\alpha\beta}F'^{\delta\alpha\beta}&=&\frac{1}{4}F_{\lambda\alpha\beta}F^{\delta\alpha\beta}-[l^\delta (\del_\lambda\del_\beta\Phi) +l_\lambda(\del^\delta \del_\beta\Phi)]k^\beta \nn \\
&&+\frac{1}{2}(\del_\beta\Phi)(\del^\beta\Phi)k_\lambda k^\delta +\Phi k^\delta k^\alpha (\del_\lambda\del_\alpha\Phi)-\Phi l_\lambda \square k^\delta.
\eea
From this expression we easily see that $F'_{\lambda\alpha\beta}F'^{\lambda\alpha\beta} = F_{\lambda\alpha\beta}F^{\lambda\alpha\beta} =0$. Moreover, 
\bea
\frac{1}{4}g'_{\nu\delta} F'_{\lambda\alpha\beta}F'^{\delta\alpha\beta}&=&\frac{1}{4}(g_{\nu\delta}+\Phi S_{\nu\delta}) F'_{\lambda\alpha\beta}F'^{\delta\alpha\beta}\\
&=&\frac{1}{4}F_{\lambda\alpha\beta}{F_\nu}^{\alpha\beta}-[l_\nu (\del_\lambda\del_\mu \Phi) +l_\lambda(\del_\nu \del_\mu \Phi)]k^\mu+\frac{1}{2}(\del_\rho\Phi)(\del^\rho\Phi)k_\lambda k_\nu \nn \\
&& -\Phi l_\lambda \square k_\nu-\Phi l_\nu \square k_\lambda+\Phi^2 l_\lambda l_\nu (\del^\alpha k_\delta)(\del_\alpha k^\delta),
\eea
where we have used the identities
\begin{eqnarray}
 F_{\lambda\alpha\beta}n^{\alpha\beta}&=&4\square k_\lambda,\\
 S_{\nu\delta}\square k^\delta &=&l_\nu k_\delta\square k^\delta.
\end{eqnarray}
We see that the right hand side matches with the left hand side.

\subsection{Matter field equations}
   \label{sec:matter}
The matter field equations are
\begin{eqnarray}
\del_\mu F^{\mu\nu\rho}&=&0.
\end{eqnarray}
Under the deformation the left hand side of this equation changes as
\begin{eqnarray}
\del'_\mu F'^{\mu\nu\rho}&=& \del_\mu F'^{\mu\nu\rho}+\Omega^\mu_{\mu\sigma}F'^{\sigma\nu\rho}+\Omega^\nu_{\mu\sigma}F'^{\mu\sigma\rho}+\Omega^\rho_{\mu\sigma}F'^{\mu\nu\sigma} \\
&=&\del_\mu F'^{\mu\nu\rho}\\
& =& \del_\mu F^{\mu\nu\rho}-\del_\mu Q^{\mu\nu\rho}. \label{matter_simp}
\end{eqnarray}
The first term in equation \eqref{matter_simp} is just the field equations for the background configuration, which is zero. For the second term in \eqref{matter_simp}, we have via \eqref{Q}
\begin{eqnarray}
Q^{\mu\nu\rho}&=&g^{\mu\sigma}g^{\nu\eta}g^{\rho\alpha}Q_{\sigma\eta\alpha}\\
&=&(\del^\mu\Phi)m^{\nu\rho}+(\del^\nu\Phi)m^{\rho\mu}+(\del^\rho\Phi)m^{\mu\nu}.
\end{eqnarray}
Applying the covariant $\del_\mu$ on this expression we find, 
\begin{eqnarray}
\nonumber
\del_\mu Q^{\mu\nu\rho}&=&(\Box\Phi) m^{\nu\rho}+(\del^\mu\Phi)[l^\rho(\del_\mu k^\nu)-l^\nu(\del_\mu k^\rho)]+(\del_\mu\del^\nu\Phi)(k^\rho l^\mu - k^\mu l^\rho)\\
&+&(\del_\mu\del^\rho\Phi)(k^\mu l^\nu- k^\nu l^\mu ).
\end{eqnarray}
Using
\begin{eqnarray}
\Box\Phi &=&0,\\
l^\mu(\del_\mu\del^\nu\Phi)&=&0,\\
k^\mu(\del_\mu\del^\nu\Phi)&=&(\del^\mu\Phi)(\del_\mu k^\nu),
\end{eqnarray}
we get
\begin{eqnarray}
\del'_\mu F'^{\mu\nu\rho}&=&\del_\mu Q^{\mu\nu\rho}~=~0.
\end{eqnarray}
Hence the matter field equations are also satisfied by the transformed configuration.

 We have shown that under the generalised Garfinkle-Vachaspati transform, solutions of IIB theory are mapped to solutions of IIB theory.

\section{BW and GMR formalisms}
\label{sec:dualities_details}
 
In this appendix, after a brief review of  the Gutowski-Martelli-Reall (GMR) and  the Bena-Warner (BW) formalisms we  relate the two notations. Similar computations were also done in \cite{Bena:2008wt, Bena:2008dw, Saxena:2005uk}.

\subsection{Gutowski-Martelli-Reall formalism}
In the GMR formalism \cite{GMR}, we work with minimal six-dimensional 
supergravity. 
We follow the notation of appendix A of 
reference \cite{Lunin:2012gp}.  The bosonic part of this theory consists of metric $g_{\mu\nu}$ and a self-dual three-form $G_{\mu\nu\rho}$. GMR showed that the metric for any supersymmetric solution of minimal 6D supergravity 
can be written as
\begin{equation}
ds^2 = -H^{-1}(dv +\beta)\left(du + \omega + \frac{\cF}{2}(dv + \beta)\right) + H h_{mn}dx^m dx^n, \label{GMR}
\end{equation}
where $h_{mn}$ is a metric on a four-dimensional almost hyper-K\"ahler base manifold, $\beta$ and $\omega$ are one-forms on this base space, while $\cF$ and $H$ are functions on the base space.

In general, the above metric only has
\be
k = \frac{\partial}{\partial u},
\ee
as the null Killing vector, i.e.,  $h_{mn}$, $\beta$, $\omega$ $\cF$ and $H$ can be $v$-dependent.  However, to compare with the Bena-Warner formalism \cite{Bena:2005va}, we must restrict to $v$-\emph{independent} solutions. For this case, 
the six-dimensional field strength $G$ takes the form
\bea
F \ = \ 2G &=& \star dH - H^{-1}(dv + \beta) \wedge\left(\frac{d\omega -\star d\omega}{2}\right)\nn \\
 &&  + H^{-1}\left(du + \omega + \frac{\cF}{2}(dv + \beta)\right) \wedge \left(d\beta +H^{-1} (dv + \beta)\wedge dH \right).
 \label{GMR_field_strength}
\eea

A detailed analysis of the Killing spinor equations shows that the equations of motion then reduce to 
 \bea
\star d\star d\cF - \frac{1}{2}(\mathcal{G}^{+})^2 &=&0, \label{GMR_EQ_1} \\
d\star dH + \frac{d\beta \wedge \mathcal{G}^{+} }{2}&=&0,  \label{GMR_EQ_2}\\
d\beta - \star d\beta &=& 0,  \label{GMR_EQ_3}\\
d\mathcal{G}^{+} &=&0.  \label{GMR_EQ_4}
\eea
In these equations, the Hodge star is with respect to $4$-dimensional base metric $h_{\mu\nu}$ and self-dual two-form $\mathcal{G}^{+}$ is defined as
\begin{equation}
\mathcal{G}^{+} = \frac{1}{2H} \left(d\omega + \star d\omega + \cF d\beta\right).
\end{equation}
We also  note that $\star d\star d\cF = -\nabla^2 \cF$  and $(\mathcal{G}^{+})^2 = (\mathcal{G}^{+})^{mn}(\mathcal{G}^{+})_{mn}$.

\subsection{Bena-Warner formalism}
Bena and Warner \cite{Bena:2005va} showed that  solutions preserving same supersymmetries as 
those of three charge black holes and black ring can be written in a general form with one forms defined on a four dimensional hyper-K\"ahler base space. Their formalism is simplest and most symmetric in the 
M-theory form, with branes intersecting on the six-torus with coordinates $(z_1, \ldots,z_6)$ as M2(12)--M2(34)--M2(56). We refer the reader to the review \cite{Bena:2007kg} for further details on brane-intersection. The metric in eleven-dimensions takes the following symmetrical form,
\be
ds_{11}^2 = ds_5^2+ ds^2_{\mathrm{T}^6},
\ee
where $ds^2_{\mathrm{T}^6}$ is metric on the six-torus, 
\be
ds^2_{\mathrm{T}^6} = (Z_2 Z_3 Z_1^{-2})^{\frac{1}{3}}(dz_1^2+dz_2^2)
+(Z_1Z_3Z_2^{-2})^{\frac{1}{3}}(dz_3^2+dz_4^2)+(Z_1Z_2Z_3^{-2})^{\frac{1}{3}}(dz_5^2+dz_6^2),
\ee
and $ds_5^2$ is the metric on five-dimensional transverse spacetime, 
\be
ds_5^2=-(Z_1Z_2Z_3)^{-\frac{2}{3}}(dt+\k)^2+(Z_1Z_2Z_3)^{\frac{1}{3}}h_{mn}dx^m dx^n, 
\ee
where $h_{mn}$ is the metric on a 4-dimensional hyper-K\"{a}hler base space.

The M-theory three-form potential $\mathcal{A}$ for this class of solutions can be written in terms of three one-form potentials $A^{(I)}$ on the five-dimensional spacetime,
\be
\mathcal{A}=A^{(1)}\wedge dz_1\wedge dz_2+A^{(2)}\wedge dz_3\wedge dz_4+A^{(3)}\wedge dz_5\wedge dz_6,
\ee
which in turn take the form,
\be
A^{(I)}  = -\frac{(dt+\k)}{Z_I}+\omega_I , \label{A_I}
\ee
where $\k$ and $\omega_I$ are one-forms on the four-dimensional base space while $Z_I$ are functions on the base space. These functions and one-forms are determined by the BW equations \cite{Bena:2005va}:
\begin{eqnarray}
d\omega_{I} &=& \star  d\omega_I, \label{BW_EQ_1}\\
d\k + \star  d\k &=& Z_I d\omega_I, \label{BW_EQ_2} \\
\nabla^2 Z_I &=& \frac{1}{2}|\epsilon_{IJK} | \star  (d\omega_J \wedge d\omega_K),\label{BW_EQ_3}
\end{eqnarray}
where the Hodge star is with respect to the four-dimensional base metric $h_{mn}$.

To compare with the GMR formalism, we convert from the M-theory form to the type IIB D1-D5-P form  using dualities and dimensional reduction (later we will truncate to six-dimensional minimal supergravity). 
Performing a dimensional reduction along the $z_6$-direction we can go from M-theory to type-IIA theory with the metric of a D2(12)--D2(34)--F1(5) brane intersection. The resulting IIA metric in the string frame is,
\begin{eqnarray}
ds_{10}^2=&-&\frac{1}{Z_3\sqrt{Z_1Z_2}} (dt+\k)^2+\sqrt{Z_1Z_2}h_{mn}dx^m dx^n \nn \\
&+&\sqrt{\frac{Z_2}{Z_1}}(dz_1^2+dz_2^2)+\sqrt{\frac{Z_1}{Z_2}}(dz_3^2+dz_4^2)+\frac{\sqrt{Z_1Z_2}}{Z_3}dz_5^2,
\end{eqnarray}
with IIA dilaton,
\be
e^{2\phi}=\frac{\sqrt{Z_1Z_2}}{Z_3},
\ee
and with three-form RR field, 
\begin{eqnarray}
C_{\mu z_1 z_2}&=&A^{(1)}_\mu, \\ C_{\mu z_3 z_4}&=&A^{(2)}_\mu,
\eea
and two-form NS-NS B-field,
\bea
 B_{\mu z_5}&=&A^{(3)}_\mu.  
\end{eqnarray}

Next we need to perform T-dualities along $z_3,z_4$ and $z_5$ directions to get D5(12345)--D1(5)--P(5) system. We recall the T-duality rules for a duality along $z$-direction:
\begin{eqnarray}
G'_{zz}&=&\frac{1}{G_{zz}},\\
G'_{\mu z}&=&\frac{B_{\mu z}}{G_{zz}}, \\
G'_{\mu\nu}&=&G_{\mu\nu}-\frac{G_{\mu z}G_{\nu z}-B_{\mu z}B_{\nu z}}{G_{zz}},\\
B'_{\mu z}&=&\frac{G_{\mu z}}{G_{zz}},\\
B'_{\mu\nu}&=&B_{\mu\nu}-\frac{B_{\mu z}G_{\nu z}-G_{\mu z}B_{\nu z}}{G_{zz}}, \\
e^{2 \phi'} &=& \frac{e^{2\phi}}{G_{zz}}, \\
{C'}^{(n)}_{\mu \ldots\nu\alpha z}&=&{C}^{(n-1)}_{\mu \ldots\nu\alpha}-(n-1)\frac{{C}^{(n-1)}_{[\mu \ldots\nu| z}G_{|\alpha]z}}{G_{zz}}, \\
C'^{(n)}_{\mu\ldots\nu\alpha\beta}&=&C^{(n+1)}_{\mu \ldots\nu\alpha\beta z}+nC^{(n-1)}_{[\mu\ldots\nu\alpha} B_{\beta]z}+n (n-1)\frac{{C}^{(n-1)}_{[\mu \ldots\nu| z}B_{|\alpha|z}G_{|\beta]z}}{G_{zz}}.
\end{eqnarray}

We perform the required dualities in two steps. Performing T-dualities along $z_3,z_4$ directions we get the following fields:
\begin{eqnarray}
ds_{10}^2=&-&\frac{1}{Z_3\sqrt{Z_1Z_2}}(dt+\k)^2+\sqrt{Z_1Z_2}h_{mn}dx^m dx^n \nn \\
&+&\sqrt{\frac{Z_2}{Z_1}}(dz_1^2+dz_2^2+dz_3^2+dz_4^2)+\frac{\sqrt{Z_1Z_2}}{Z_3}dz_5^2,
\end{eqnarray}
\be
e^{2\phi} = \frac{Z_2^{3/2}}{Z_3\sqrt{Z_1}},
\ee
\begin{align}
C^{(5)}_{\mu z_1 z_2 z_3 z_4}&= A^{(1)}_\mu, & 
C^{(1)}_\mu &=  -A^{(2)}_\mu , &
B_{\mu z_5}&=A^{(3)}_\mu.
\end{align}

Now doing T-duality along $z_5$-direction, we get our required D1-D5-P configuration. The IIB dilaton reads:
\be
e^{2\phi}=\frac{Z_2}{Z_1},
\ee 
and  the metric takes the form,
\begin{eqnarray}
ds_{10}^2&=&-\frac{1}{Z_3\sqrt{Z_1Z_2}}(dt+\k)^2+\sqrt{Z_1Z_2}h_{mn}dx^m dx^n \nn \\ 
& & +\frac{Z_3}{\sqrt{Z_1Z_2}}(dz_5+A^{(3)}_\mu dx^\mu )^2+\sqrt{\frac{Z_2}{Z_1}}(dz_1^2 +dz_2^2+ dz_3^2 + d z_4^2), \label{metric_after_T_dualities}
\end{eqnarray}
together with the associated RR-field components,
\begin{eqnarray}
C^{(6)}&=&A^{(1)}_\mu dx^\mu \wedge dx^1\wedge dx^2\wedge dx^3\wedge dx^4\wedge dx^5 + A^{(1)}_\mu A^{(3)}_\nu dx^\mu \wedge  dx^\nu \wedge dx^1\wedge dx^2\wedge dx^3\wedge dx^4,  \nonumber \\ 
C^{(2)}&=&-A^{(2)}_\mu dx^\mu \wedge dx^5- A^{(2)}_\mu A^{(3)}_\nu dx^\mu \wedge  dx^\nu.
\end{eqnarray}
We can dualize the $6$-form potential to get a $2$-form potential. This is a tedious step. Fortunately, we do not need to do this electromagnetic duality. Comparing metric \eqref{metric_after_T_dualities} to the GMR form, we obtain a complete dictionary between the GMR and the BW variables. Using this dictionary we can convert the GMR form of the field strength \eqref{GMR_field_strength} into  the BW variables. We expect the electromagnetic duality to give the same result.

Since GMR formalism is for minimal six-dimensional supergravity, in order to compare the above configuration with the GMR form we must set $Z_1=Z_2$. In that case, the dilaton vanishes $e^{2\phi}=1$. 
Inserting  
 $A^{(3)}_\mu dx^\mu $ from  \eqref{A_I} in  metric  \eqref{metric_after_T_dualities} we get, 
 \be
ds_{10}^2=- 2 Z_1^{-1}(dt + \k)(dz_5 +\omega_3)  + Z_3 Z_1^{-1}(d z_5+\omega_3 )^2+Z_1 h_{mn}dx^m dx^n  + ds_{\mathrm{T}_4}^2,
\ee
where
\be
 ds_{\mathrm{T}_4}^2 = dz_1^2 +dz_2^2+ dz_3^2 + d z_4^2,
\ee
is the metric on the four-torus.
To match with the GMR form \eqref{GMR}, we identify
\begin{align}
&  z_5 ~=~v,  &
& Z_1 ~=~H, \nn \\ 
&  Z_3 ~=~1-\frac{\cF}{2} ,&
& \omega_3 ~=~ \beta, \nn \\
& \k~=~\frac{\beta + \omega}{2} , & 
& t~=~ \frac{u+v}{2} . \label{dictionary}
\end{align}

Using the identification \eqref{dictionary} in the GMR field strength \eqref{GMR_field_strength}, we get
\begin{eqnarray}
G&=&\frac{1}{2} \star  dZ_1-\frac{1}{4Z_1} (dz_5 +\omega_3)\wedge [d\k-\star  d\k]+\frac{1}{2Z_1}[(dt+\k)-\frac{Z_3}{2}(dz_5+\omega_3)]\wedge d\omega_3 \nn \\
&&-\frac{1}{2Z_1^2}(dz_5+\omega_3)\wedge(dt+\k) \wedge dZ_1,
\end{eqnarray}
 which using the BW equations of motion simplifies to 
 \begin{eqnarray}
2G&=&\ \star  dZ_1 + d \left[(dz_5+\omega_3)]\wedge \left(\frac{dt +\k}{Z_1} -\omega_1 \right)\right] 
+ \omega_1 \wedge d\omega_3 .
\end{eqnarray} 
The RR field strength in ten dimensions is normalised as $F = 2G$, with the associated 2-form field
\be
C=-\left[\left(\frac{dt+\k}{Z_1}-\omega_1\right)\wedge (dz_5+\omega_3)\right]+ \sigma \label{2formpotential},
\ee
where an explicit expression for $\sigma$ cannot be obtained in general. It satisfies,
\be
d\sigma = \star dZ_1+\omega_1\wedge d\omega_3. \label{sigma_eq}
\ee 
One can easily check that the three form $ \star dZ_1+\omega_1\wedge d\omega_3 $ appearing on the right hand side of equation \eqref{sigma_eq} is exact due to BW equations of motion for $Z_1$.

\subsection{Relation between GMR and BW }

Now that we have a simple dictionary \eqref{dictionary} we can easily  relate BW and GMR equations of motion. On the GMR side, we look at $v$-independent solutions while on the  BW side we consider solutions with  $Z_1 =Z_2$ and $\omega_1 =\omega_2$.

We consider BW equations and using the dictionary transform them into  GMR  equations. 
Consider BW equation \eqref{BW_EQ_2},
\be
d\k + \star  d\k  = 2  Z_1 d\omega_1 + Z_3d\omega_3.
\ee
Rewriting this equation using dictionary \eqref{dictionary}, we have
\bea
2  d\omega_1 &=& \frac{1}{Z_1}\left(d\k + \star  d\k - Z_3d\omega_3\right) \\ 
&=& \frac{1}{2H}\left(d\omega + \star  d\omega +2 (1 - Z_3)d\beta \right) = \frac{1}{H}\left( d\omega + \star  d\omega + \cF d\beta \right) = \mathcal{G}^+,
\eea
where we have used the fact that $d\beta =d\omega_3$ is self dual, cf.~\eqref{BW_EQ_1}. It then immediately follows that  $d\mathcal{G}^+ =0$, which is 
one of the GMR equations, cf.~\eqref{GMR_EQ_4}. 
Similarly, from the BW  scalar equations  \eqref{BW_EQ_3} for $Z_1$ we have,
\begin{equation}
\nabla^2 Z_1 =\nabla^2 H = - \star  d \star  d H = \star (d\omega_3 \wedge d\omega_2) = \star  \left(\frac{d\beta \wedge \mathcal{G}^{+} }{2}\right),
\end{equation}
which implies \eqref{GMR_EQ_2}. Similarly,
\begin{equation}
\nabla^2 Z_3 =-\frac{1}{2}\nabla^2 \cF = \star (d\omega_1 \wedge d\omega_2) =  \star  \left(\frac{\mathcal{G}^+ \wedge \mathcal{G}^{+} }{4}\right),
\end{equation}
which implies \eqref{GMR_EQ_1}.


\begin{thebibliography}{99}
   
   
\bibitem{Sen:1995in} 
  A.~Sen,
  ``Extremal black holes and elementary string states,''
  Mod.\ Phys.\ Lett.\ A {\bf 10}, 2081 (1995)
  doi:10.1142/S0217732395002234
  [hep-th/9504147].
  
  
  
\bibitem{Strominger:1996sh} 
  A.~Strominger and C.~Vafa,
  ``Microscopic origin of the Bekenstein-Hawking entropy,''
  Phys.\ Lett.\ B {\bf 379}, 99 (1996)
  doi:10.1016/0370-2693(96)00345-0
  [hep-th/9601029].

  
\bibitem{David:2002wn} 
  J.~R.~David, G.~Mandal and S.~R.~Wadia,
  ``Microscopic formulation of black holes in string theory,''
  Phys.\ Rept.\  {\bf 369}, 549 (2002)
  doi:10.1016/S0370-1573(02)00271-5
  [hep-th/0203048].
  
  
\bibitem{Mathur:2005ai} 
  S.~D.~Mathur,
  ``The Quantum structure of black holes,''
  Class.\ Quant.\ Grav.\  {\bf 23}, R115 (2006)
  doi:10.1088/0264-9381/23/11/R01
  [hep-th/0510180].
  

    
\bibitem{Sen:2007qy} 
  A.~Sen,
  ``Black Hole Entropy Function, Attractors and Precision Counting of Microstates,''
  Gen.\ Rel.\ Grav.\  {\bf 40}, 2249 (2008)
  doi:10.1007/s10714-008-0626-4
  [arXiv:0708.1270 [hep-th]].
  
  

  
\bibitem{Mandal:2010cj} 
  I.~Mandal and A.~Sen,
  ``Black Hole Microstate Counting and its Macroscopic Counterpart,''
  Nucl.\ Phys.\ Proc.\ Suppl.\  {\bf 216}, 147 (2011)
  [Class.\ Quant.\ Grav.\  {\bf 27}, 214003 (2010)]
  doi:10.1088/0264-9381/27/21/214003
  [arXiv:1008.3801 [hep-th]].
  
\bibitem{Dabholkar:2012zz} 
  A.~Dabholkar and S.~Nampuri,
  ``Quantum black holes,''
  Lect.\ Notes Phys.\  {\bf 851}, 165 (2012)
  doi:10.1007/978-3-642-25947-05
  [arXiv:1208.4814 [hep-th]].



 
\bibitem{Mathur:2005zp} 
  S.~D.~Mathur,
  ``The Fuzzball proposal for black holes: An Elementary review,''
  Fortsch.\ Phys.\  {\bf 53}, 793 (2005)
  doi:10.1002/prop.200410203
  [hep-th/0502050].
  
  
\bibitem{Bena:2007kg} 
  I.~Bena and N.~P.~Warner,
  ``Black holes, black rings and their microstates,''
  Lect.\ Notes Phys.\  {\bf 755}, 1 (2008)
    doi:10.1007/978-3-540-79523-0-1
  [hep-th/0701216].


\bibitem{Skenderis:2008qn} 
  K.~Skenderis and M.~Taylor,
  ``The fuzzball proposal for black holes,''
  Phys.\ Rept.\  {\bf 467}, 117 (2008)
  doi:10.1016/j.physrep.2008.08.001
  [arXiv:0804.0552 [hep-th]].
  
  
\bibitem{Chowdhury:2010ct} 
  B.~D.~Chowdhury and A.~Virmani,
  ``Modave Lectures on Fuzzballs and Emission from the D1-D5 System,''
  arXiv:1001.1444 [hep-th].




\bibitem{Tod:1983pm} 
  K.~P.~Tod,
  ``All Metrics Admitting Supercovariantly Constant Spinors,''
  Phys.\ Lett.\  {\bf 121B}, 241 (1983).
  doi:10.1016/0370-2693(83)90797-9
  

\bibitem{Gauntlett:2002nw} 
  J.~P.~Gauntlett, J.~B.~Gutowski, C.~M.~Hull, S.~Pakis and H.~S.~Reall,
  ``All supersymmetric solutions of minimal supergravity in five- dimensions,''
  Class.\ Quant.\ Grav.\  {\bf 20}, 4587 (2003)
  doi:10.1088/0264-9381/20/21/005
  [hep-th/0209114].

\bibitem{Gauntlett:2004qy} 
  J.~P.~Gauntlett and J.~B.~Gutowski,
  ``General concentric black rings,''
  Phys.\ Rev.\ D {\bf 71}, 045002 (2005)
  doi:10.1103/PhysRevD.71.045002
  [hep-th/0408122].
 
 
 
  
%
\bibitem{Bena:2005va} 
  I.~Bena and N.~P.~Warner,
  ``Bubbling supertubes and foaming black holes,''
  Phys.\ Rev.\ D {\bf 74}, 066001 (2006)
  doi:10.1103/PhysRevD.74.066001
  [hep-th/0505166].
  
\bibitem{Berglund:2005vb} 
  P.~Berglund, E.~G.~Gimon and T.~S.~Levi,
  ``Supergravity microstates for BPS black holes and black rings,''
  JHEP {\bf 0606}, 007 (2006)
  doi:10.1088/1126-6708/2006/06/007
  [hep-th/0505167].

 
 
\bibitem{GMR} 
  J.~B.~Gutowski, D.~Martelli and H.~S.~Reall,
  ``All Supersymmetric solutions of minimal supergravity in six- dimensions,''
  Class.\ Quant.\ Grav.\  {\bf 20}, 5049 (2003)
  doi:10.1088/0264-9381/20/23/008
  [hep-th/0306235].



\bibitem{Garfinkle:1990jq} 
  D.~Garfinkle and T.~Vachaspati,
  ``Cosmic string traveling waves,''
  Phys.\ Rev.\ D {\bf 42}, 1960 (1990).
  doi:10.1103/PhysRevD.42.1960
  
\bibitem{Kaloper:1996hr} 
  N.~Kaloper, R.~C.~Myers and H.~Roussel,
  ``Wavy strings: Black or bright?,''
  Phys.\ Rev.\ D {\bf 55}, 7625 (1997)
  doi:10.1103/PhysRevD.55.7625
  [hep-th/9612248].
  
  


\bibitem{Dabholkar:1995nc} 
  A.~Dabholkar, J.~P.~Gauntlett, J.~A.~Harvey and D.~Waldram,
  ``Strings as solitons and black holes as strings,''
  Nucl.\ Phys.\ B {\bf 474}, 85 (1996)
  doi:10.1016/0550-3213(96)00266-0
  [hep-th/9511053].



\bibitem{Horowitz:1996th} 
  G.~T.~Horowitz and D.~Marolf,
  ``Counting states of black strings with traveling waves,''
  Phys.\ Rev.\ D {\bf 55}, 835 (1997)
  doi:10.1103/PhysRevD.55.835
  [hep-th/9605224].


  
\bibitem{Banados:1999tw} 
  M.~Banados, A.~Chamblin and G.~W.~Gibbons,
  ``Branes, AdS gravitons and Virasoro symmetry,''
  Phys.\ Rev.\ D {\bf 61}, 081901 (2000)
  doi:10.1103/PhysRevD.61.081901
  [hep-th/9911101].
  
\bibitem{Hubeny:2003ug} 
  V.~E.~Hubeny and M.~Rangamani,
  ``Horizons and plane waves: A Review,''
  Mod.\ Phys.\ Lett.\ A {\bf 18}, 2699 (2003)
  doi:10.1142/S0217732303012428
  [hep-th/0311053].
  
    
\bibitem{Balasubramanian:2010ys} 
  V.~Balasubramanian, J.~Parsons and S.~F.~Ross,
  ``States of a chiral 2d CFT,''
  Class.\ Quant.\ Grav.\  {\bf 28}, 045004 (2011)
  doi:10.1088/0264-9381/28/4/045004
  [arXiv:1011.1803 [hep-th]].
  
  
\bibitem{Lunin:2012gp} 
  O.~Lunin, S.~D.~Mathur and D.~Turton,
  ``Adding momentum to supersymmetric geometries,''
  Nucl.\ Phys.\ B {\bf 868}, 383 (2013)
  doi:10.1016/j.nuclphysb.2012.11.017
  [arXiv:1208.1770 [hep-th]].
  
\bibitem{Mathur:2011gz} 
  S.~D.~Mathur and D.~Turton,
  ``Microstates at the boundary of AdS,''
  JHEP {\bf 1205}, 014 (2012)
  doi:10.1007/JHEP05(2012)014
  [arXiv:1112.6413 [hep-th]].
  
  
  \bibitem{Mathur:2012tj} 
  S.~D.~Mathur and D.~Turton,
  ``Momentum-carrying waves on D1-D5 microstate geometries,''
  Nucl.\ Phys.\ B {\bf 862}, 764 (2012)
  doi:10.1016/j.nuclphysb.2012.05.014
  [arXiv:1202.6421 [hep-th]].
  
  

\bibitem{Balasubramanian:2000rt} 
  V.~Balasubramanian, J.~de Boer, E.~Keski-Vakkuri and S.~F.~Ross,
  ``Supersymmetric conical defects: Towards a string theoretic description of black hole formation,''
  Phys.\ Rev.\ D {\bf 64}, 064011 (2001)
  doi:10.1103/PhysRevD.64.064011
  [hep-th/0011217].
  
\bibitem{Maldacena:2000dr} 
  J.~M.~Maldacena and L.~Maoz,
  ``Desingularization by rotation,''
  JHEP {\bf 0212}, 055 (2002)
  doi:10.1088/1126-6708/2002/12/055
  [hep-th/0012025].


\bibitem{Lunin:2001jy} 
  O.~Lunin and S.~D.~Mathur,
  ``AdS/CFT duality and the black hole information paradox,''
  Nucl.\ Phys.\ B {\bf 623}, 342 (2002)
  doi:10.1016/S0550-3213(01)00620-4
  [hep-th/0109154].
  
  
\bibitem{gms1} 
  S.~Giusto, S.~D.~Mathur and A.~Saxena,
  ``Dual geometries for a set of 3-charge microstates,''
  Nucl.\ Phys.\ B {\bf 701}, 357 (2004)
  doi:10.1016/j.nuclphysb.2004.09.001
  [hep-th/0405017].
  
  
\bibitem{gms2} 
  S.~Giusto, S.~D.~Mathur and A.~Saxena,
  ``3-charge geometries and their CFT duals,''
  Nucl.\ Phys.\ B {\bf 710}, 425 (2005)
  doi:10.1016/j.nuclphysb.2005.01.009
  [hep-th/0406103].


\bibitem{Jejjala:2005yu} 
  V.~Jejjala, O.~Madden, S.~F.~Ross and G.~Titchener,
  ``Non-supersymmetric smooth geometries and D1-D5-P bound states,''
  Phys.\ Rev.\ D {\bf 71}, 124030 (2005)
  doi:10.1103/PhysRevD.71.124030
  [hep-th/0504181].

\bibitem{Giusto:2012yz}
  S.~Giusto, O.~Lunin, S.~D.~Mathur and D.~Turton,
  ``D1-D5-P microstates at the cap,''
  JHEP {\bf 1302}, 050 (2013)
  doi:10.1007/JHEP02(2013)050
  [arXiv:1211.0306 [hep-th]].


\bibitem{Chakrabarty:2015foa} 
  B.~Chakrabarty, D.~Turton and A.~Virmani,
  ``Holographic description of non supersymmetric orbifolded D1-D5-P solutions,''
  JHEP {\bf 1511}, 063 (2015)
  doi:10.1007/JHEP11(2015)063
  [arXiv:1508.01231 [hep-th]].


\bibitem{Giusto:2004kj} 
  S.~Giusto and S.~D.~Mathur,
  ``Geometry of D1-D5-P bound states,''
  Nucl.\ Phys.\ B {\bf 729}, 203 (2005)
  doi:10.1016/j.nuclphysb.2005.09.037
  [hep-th/0409067].
  
\bibitem{Harmark:2004ch} 
  T.~Harmark and N.~A.~Obers,
  ``General definition of gravitational tension,''
  JHEP {\bf 0405}, 043 (2004)
  doi:10.1088/1126-6708/2004/05/043
  [hep-th/0403103].
  
  
\bibitem{Roy:2016zzv} 
  P.~Roy, Y.~K.~Srivastava and A.~Virmani,
  ``Hair on non-extremal D1-D5 bound states,''
  JHEP {\bf 1609}, 145 (2016)
  doi:10.1007/JHEP09(2016)145
  [arXiv:1607.05405 [hep-th]].
  
 \bibitem{Ett:2010by} 
  B.~Ett and D.~Kastor,
  ``An Extended Kerr-Schild Ansatz,''
  Class.\ Quant.\ Grav.\  {\bf 27}, 185024 (2010)
  doi:10.1088/0264-9381/27/18/185024
  [arXiv:1002.4378 [hep-th]].
  
\bibitem{Malek:2014dta} 
  T.~Malek 
  ``Extended Kerr-Schild spacetimes: General properties and some explicit examples,''
  Class.\ Quant.\ Grav.\  {\bf 31}, 185013 (2014)
  doi:10.1088/0264-9381/31/18/185013
  [arXiv:1401.1060 [gr-qc]].
  
  
\bibitem{Julia:1994bs} 
  B.~Julia and H.~Nicolai,
  ``Null Killing vector dimensional reduction and Galilean geometrodynamics,''
  Nucl.\ Phys.\ B {\bf 439}, 291 (1995)
  doi:10.1016/0550-3213(94)00584-2
  [hep-th/9412002].
  
  

\bibitem{Bena:2008wt} 
  I.~Bena, N.~Bobev and N.~P.~Warner,
  ``Spectral Flow, and the Spectrum of Multi-Center Solutions,''
  Phys.\ Rev.\ D {\bf 77}, 125025 (2008)
  doi:10.1103/PhysRevD.77.125025
  [arXiv:0803.1203 [hep-th]].
  
\bibitem{Bena:2008dw} 
  I.~Bena, N.~Bobev, C.~Ruef and N.~P.~Warner,
  ``Supertubes in Bubbling Backgrounds: Born-Infeld Meets Supergravity,''
  JHEP {\bf 0907}, 106 (2009)
  doi:10.1088/1126-6708/2009/07/106
  [arXiv:0812.2942 [hep-th]].
  

\bibitem{Saxena:2005uk} 
  A.~Saxena, G.~Potvin, S.~Giusto and A.~W.~Peet,
  ``Smooth geometries with four charges in four dimensions,''
  JHEP {\bf 0604}, 010 (2006)
  doi:10.1088/1126-6708/2006/04/010
  [hep-th/0509214].


\end{thebibliography}
\end{document}